\documentclass[letterpaper,oneside,final,notitlepage,onecolumn]{article}
\usepackage{graphicx}
\usepackage{amsmath}
\usepackage{amsfonts}
\usepackage{amssymb}

\begin{document}

\begin{center}
{\Large Hierarchic Theory of Condensed Matter:}

{\Large \medskip}

{\Large Role of water in protein dynamics, function \& cancer emergency}

\bigskip

{\large Alex Kaivarainen}

\bigskip

{\large JBL, University of Turku, FIN-20520, Turku, Finland}

\medskip

{\large \thinspace http://www.karelia.ru/\symbol{126}alexk}

{\large H2o@karelia.ru}

\medskip

\medskip
\end{center}

\begin{quotation}
\thinspace\thinspace\thinspace\textbf{Materials, presented in this original
article are based on following publications:}

\smallskip

\textbf{[1]. A. Kaivarainen. Book: Hierarchic Concept of Matter and Field.
Water, biosystems and elementary particles. New York, NY, 1995, ISBN
\ 0-9642557-0-7 }

\textbf{[2]. \thinspace A. Kaivarainen. New Hierarchic Theory of Matter
General for Liquids and Solids: dynamics, thermodynamics and mesoscopic
structure of water and ice (see: http://www.karelia.ru/\symbol{126}alexk
\ [New articles] ).}

\textbf{[3].} \textbf{A. Kaivarainen. Hierarchic Concept of Condensed Matter
and its Interaction with Light: New Theories of Light Refraction, Brillouin
Scattering\ and M\"{o}ssbauer effect (see: http://www.karelia.ru/\symbol{126}%
alexk \ [New articles]). }

\textbf{[4]. A. Kaivarainen. Hierarchic Concept of Condensed Matter :
Interrelation between mesoscopic and macroscopic properties (see:
http://www.karelia.ru/\symbol{126}alexk \ [New articles]). }

\textbf{[5]. A. Kaivarainen. Hierarchic Theory of Complex Systems (see URL:
http://www.karelia.ru/\symbol{126}alexk \ [New articles]). }

\medskip\ \ \ \ \textbf{See also papers at Los Alamos archives at: \ }{\large http://arXiv.org/find/physics/1/au:+Kaivarainen\_A/0/1/0/all/0/1}

\textbf{CONTENTS\ OF ARTICLE:}

\smallskip
\end{quotation}

\textbf{Introduction to new Hierarchic theory of condensed matter}

\textbf{1. Role of inter-domain water clusters in large-scale dynamics of
proteins }

\textbf{2. Description of large-scale dynamics of proteins, based on
generalized. Stokes-Einstein and Eyring-Polany equation}

\textbf{3. Dynamic model of protein-ligand complexes formation}

\textbf{4. The life-time of quasiparticles and frequencies of their excitation}

\textbf{5. Mesoscopic mechanism of enzyme catalysis }

\textbf{6. The mechanism of ATP hydrolysis energy utilization in muscle
contraction and protein polymerization}

\textbf{7. Water activity as a regulative factor in the intra- and inter-cell processes}

\textbf{8. Water and cancer}

\medskip

\begin{quotation}
\textbf{Computerized verification of described here new models and theories
has been presented, using special computer program, based on our new
Hierarchic Theory of Condensed Matter (copyright, 1997, A. Kaivarainen).}
\end{quotation}

\bigskip

=====================================================================\bigskip

\begin{center}
{\large Introduction to new}

{\large Hierarchic theory of condensed matter (http://arXiv.org/abs/physics/00030044)}
\end{center}

{\large \smallskip}

\textbf{\ A basically new hierarchic quantitative theory, general for solids
and liquids, has been developed.}

\textbf{It is assumed, that unharmonic oscillations of particles in any
condensed matter lead to emergence of three-dimensional (3D) superposition of
standing de Broglie waves of molecules, electromagnetic and acoustic waves.
Consequently, any condensed matter could be considered as a gas of 3D standing
waves of corresponding nature. Our approach unifies and develops strongly the
Einstein's and Debye's models.}

\ \textbf{Collective excitations, like 3D standing de Broglie waves of
molecules, representing at certain conditions the mesoscopic molecular Bose
condensate, were analyzed, as a background of hierarchic model of condensed matter.}

\smallskip

\textbf{The most probable de Broglie wave (wave B) length is determined by the
ratio of Plank constant to the most probable impulse of molecules, or by ratio
of its most probable phase velocity to frequency. The waves B are related to
molecular translations (tr) and librations (lb).}

As the quantum dynamics of condensed matter does not follow in general case
the classical Maxwell-Boltzmann distribution, the real most probable de
Broglie wave length can exceed the classical thermal de Broglie wave length
and the distance between centers of molecules many times.

\textit{This makes possible the atomic and molecular Bose condensation in
solids and liquids at temperatures, below boiling point. It is one of the most
important results of new theory, which we have confirmed by computer
simulations on examples of water and ice.}

\smallskip

\textbf{Four strongly interrelated }new types of quasiparticles (collective
excitations) were introduced in our hierarchic model:

1.~\textit{Effectons (tr and lb)}, existing in "acoustic" (a) and "optic" (b)
states represent the coherent clusters in general case\textbf{; }

2.~\textit{Convertons}, corresponding to interconversions between \textit{tr
}and \textit{lb }types of the effectons (flickering clusters);

3.~\textit{Transitons} are the intermediate $\left[  a\rightleftharpoons
b\right]  $ transition states of the \textit{tr} and \textit{lb} effectons;

4.~\textit{Deformons} are the 3D superposition of IR electromagnetic or
acoustic waves, activated by \textit{transitons }and \textit{convertons. }\smallskip

\smallskip

\ \textbf{Primary effectons }(\textit{tr and lb) }are formed by 3D
superposition of the \textbf{most probable standing de Broglie waves }of the
oscillating ions, atoms or molecules. The volume of effectons (tr and lb) may
contain from less than one, to tens and even thousands of molecules. The first
condition means validity of \textbf{classical }approximation in description of
the subsystems of the effectons. The second one points to \textbf{quantum
properties} \textbf{of coherent clusters due to molecular Bose condensation}%
\textit{. }

\ The liquids are semiclassical systems because their primary (tr) effectons
contain less than one molecule and primary (lb) effectons - more than one
molecule. \textit{The solids are quantum systems totally because both kind of
their primary effectons (tr and lb) are molecular Bose condensates.}%
\textbf{\ These consequences of our theory are confirmed by computer
calculations. }

\ The 1st order $\left[  gas\rightarrow\,liquid\right]  $ transition is
accompanied by strong decreasing of rotational (librational) degrees of
freedom due to emergence of primary (lb) effectons and $\left[
liquid\rightarrow\,solid\right]  $ transition - by decreasing of translational
degrees of freedom due to Bose-condensation of primary (tr) effectons.

\ \textbf{In the general case the effecton can be approximated by
parallelepiped with edges corresponding to de Broglie waves length in three
selected directions (1, 2, 3), related to the symmetry of the molecular
dynamics. In the case of isotropic molecular motion the effectons' shape may
be approximated by cube.}

\textbf{The edge-length of primary effectons (tr and lb) can be considered as
the ''parameter of order''.}

\smallskip

The in-phase oscillations of molecules in the effectons correspond to the
effecton's (a) - \textit{acoustic }state and the counterphase oscillations
correspond to their (b) - \textit{optic }state. States (a) and (b) of the
effectons differ in potential energy only, however, their kinetic energies,
impulses and spatial dimensions - are the same. The \textit{b}-state of the
effectons has a common feature with \textbf{Fr\"{o}lich's polar mode. }

\smallskip

\textbf{The }$(a\rightarrow b)$\textbf{\ or }$(b\rightarrow a)$%
\textbf{\ transition states of the primary effectons (tr and lb), defined
as\ primary transitons, are accompanied by a change in molecule polarizability
and dipole moment without density fluctuations. At this case they lead to
absorption or radiation of IR photons, respectively.}

\textbf{\ Superposition (interception) of three internal standing IR photons
of different directions (1,2,3) - forms primary electromagnetic deformons (tr
and lb).}

\ On the other hand, the [lb$\rightleftharpoons\,$tr] \textit{convertons }and
\textit{secondary transitons} are accompanied by the density fluctuations,
leading to \textit{absorption or radiation of phonons}.

\textit{Superposition resulting from interception} of standing phonons in
three directions (1,2,3), forms \textbf{secondary acoustic deformons (tr and
lb). }

\smallskip

\ \textit{Correlated collective excitations }of primary and secondary
effectons and deformons (tr and lb)\textbf{, }localized in the volume of
primary \textit{tr }and \textit{lb electromagnetic }deformons\textbf{, }lead
to origination of \textbf{macroeffectons, macrotransitons}\textit{\ }and
\textbf{macrodeformons }(tr and lb respectively)\textbf{. }

\ \textit{Correlated simultaneous excitations of \thinspace tr and lb}
\textit{macroeffectons }in the volume of superimposed \textit{tr }and
\textit{lb }electromagnetic deformons lead to origination of
\textbf{supereffectons. }

\ In turn, the coherent excitation of \textit{both: tr }and \textit{lb
macrodeformons and macroconvertons }in the same volume means creation of
\textbf{superdeformons.} Superdeformons are the biggest (cavitational)
fluctuations, leading to microbubbles in liquids and to local defects in solids.

\smallskip

\ \textbf{Total number of quasiparticles of condensed matter equal to 4!=24,
reflects all of possible combinations of the four basic ones [1-4], introduced
above. This set of collective excitations in the form of ''gas'' of 3D
standing waves of three types: de Broglie, acoustic and electromagnetic - is
shown to be able to explain virtually all the properties of all condensed matter.}

\ \textit{The important positive feature of our hierarchic model of matter is
that it does not need the semi-empiric intermolecular potentials for
calculations, which are unavoidable in existing theories of many body systems.
The potential energy of intermolecular interaction is involved indirectly in
dimensions and stability of quasiparticles, introduced in our model.}

{\large \ The main formulae of theory are the same for liquids and solids and
include following experimental parameters, which take into account their
different properties:}

$\left[  1\right]  $\textbf{- Positions of (tr) and (lb) bands in oscillatory spectra;}

$\left[  2\right]  $\textbf{- Sound velocity; }$\,$

$\left[  3\right]  $\textbf{- Density; }

$\left[  4\right]  $\textbf{- Refraction index (extrapolated to the infinitive
wave length of photon}$)$\textbf{.}

\textit{\ The knowledge of these four basic parameters at the same temperature
and pressure makes it possible using our computer program, to evaluate more
than 300 important characteristics of any condensed matter. Among them are
such as: total internal energy, kinetic and potential energies, heat-capacity
and thermal conductivity, surface tension, vapor pressure, viscosity,
coefficient of self-diffusion, osmotic pressure, solvent activity, etc. Most
of calculated parameters are hidden, i.e. inaccessible to direct experimental measurement.}

\ The new interpretation and evaluation of Brillouin light scattering and
M\"{o}ssbauer effect parameters may also be done on the basis of hierarchic
theory. Mesoscopic scenarios of turbulence, superconductivity and superfluity
are elaborated.

\ Some original aspects of water in organization and large-scale dynamics of
biosystems - such as proteins, DNA, microtubules, membranes and regulative
role of water in cytoplasm, cancer development, quantum neurodynamics, etc.
have been analyzed in the framework of Hierarchic theory.

\medskip

\textbf{Computerized verification of our Hierarchic concept of matter on
examples of water and ice is performed, using special computer program:
Comprehensive Analyzer of Matter Properties (CAMP, copyright, 1997,
Kaivarainen). The new optoacoustic device, based on this program, with
possibilities much wider, than that of IR, Raman and Brillouin spectrometers,
has been proposed (\thinspace http://www.karelia.ru/\symbol{126}alexk \ \ [CAMP]).}

\smallskip

\textbf{This is the first theory able to predict all known experimental
temperature anomalies for water and ice. The conformity between theory and
experiment is very good even without any adjustable parameters. }

\textbf{The hierarchic concept creates a bridge between micro- and macro-
phenomena, dynamics and thermodynamics, liquids and solids in terms of quantum
physics. }

\bigskip

\begin{center}
===============================================================\smallskip

\bigskip

{\large 1. Role of inter-domain water clusters in large-scale dynamics of proteins}
\end{center}

\smallskip

\textbf{The functioning of proteins, namely antibodies, enzymes, is caused by
the physicochemical properties, geometry and dynamics of their active sites.
The mobility of an active site is related to the dynamics of the residual part
of a protein molecule, its hydration shell and the properties of a free solvent.}

\textbf{The dynamic model of a protein proposed in 1975 and supported nowadays
with numerous data (K\"{a}iv\"{a}r\"{a}inen, }$1985,1989b)$\textbf{, is based
on the following statements:}

\textbf{1.~A protein molecule contains one or more cavities or clefts capable
to large scale fluctuations - pulsations between two states: ''closed'' (A)
and ''open'' }$($\textbf{B) with lesser and bigger accessibility to water. }

\textbf{The frequency of pulsations }$\left(  \nu_{A\Leftrightarrow B}\right)
$\textbf{:}%

\[
10^{4}s^{-1}\le\nu_{A\Leftrightarrow B}\le10^{7}s^{-1}
\]
\textbf{depends on the structure of protein, its ligand state, temperature and
solvent viscosity. Transitions between A and B states are the result of the
relative displacements of protein domains and subunits forming the cavities;}

\textbf{2.~The water, interacting with protein, consists of two main fractions.}

\textbf{The 1st major fraction, which solvates the outer surface regions of
protein has less apparent cooperative properties than the 2nd minor fraction
confined to ''open'' cavities. Water molecules, interacting with the cavity in
the ''open'' (B)-state form a cooperative cluster, whose lifetime }%
$(\ge10^{-10}s^{-1}).\;$\textbf{Properties of clusters are determined by the
geometry, mobility and polarity of the cavity, as well as by temperature and pressure.}

\medskip

It is seen from X-ray structural data that the protein cavities: active sites
(AS), other interdomain clefts, the space between subunits of oligomeric
proteins, have a high nonpolar residues content. In contrast to the small
intra domain holes isolated from the outer medium, which sometimes contain
several $H_{2}O$ molecules, the interdomain and intersubunit cavities can
contain several dozens of molecules (Fig. 1), exchanging with bulk water.

\smallskip

{\large The development of the above dynamic model has lead us to the
following classification of dynamics in the native globular proteins.}

\textbf{1. Small-scale (SS) dynamics: }

low amplitude ($\le$ 1 \AA) thermal fluctuations of atoms, aminoacids
residues, and displacements of $\alpha$-helixes and $\beta$-structures within
domains and subunits, at \textbf{which the effective Stokes radius of domains
does not change.} This type of motion, related to domain stability, can differ
in the content of A and B conformers (Fig. 1, dashed line). The range of
characteristic times at SS dynamics is $(10^{-4}-10^{-11})s$, determined by
activation energies of corresponding transitions.

\textbf{2. Large-scale (LS) dynamics: }

\textbf{is subdivided into LS-pulsations and LS-librations with a character of
limited diffusion of domains and subunits of proteins:}

\textbf{LS-~pulsations }are represented by relative translational-rotational
displacements of domains and subunits at distances $\ge3\AA$. Thus, the
cavities, which are formed by domains, fluctuate between states with less (A)
and more (B) water-accessibility. The life-times of these states depending on
protein structure and external conditions are in the limits of$\;\left(
10^{-4}-10^{-7}\right)  $ s.

In accordance to our model, one of contributions to this time is determined by
frequency of excitations of $[lb/tr]$ \textbf{macroconvertons.} The frequency
of macroconvertons excitation at normal conditions is about $10^{7}\,(1/s)$.

The pulsation frequency of big multi-subunit oligomeric proteins of about
10$^{4}\,(1/s)$ could be related to stronger fluctuations of water cluster in
their central cavity like \textbf{macrodeformons} or even
\textbf{superdeformons }(Fig.3c,d).

The life-times of (A) and (B) conformer markedly exceeds the time of
transitions between them $\simeq(10^{-9}- 10^{-11})$ s.

The $\left(  A\Leftrightarrow B\right)  $ pulsations of various cavities in
proteins could be correlated. The corresponding A and B conformers have
different Stokes radii and effective volume.

The geometrical deformation of the inter-subunits large central cavity of
oligomeric proteins and the destabilization of the water cluster located in it
lead to relaxational change of $(A\Leftrightarrow B)$ equilibrium constant:%

\[
K_{A\Leftrightarrow B}=\exp\left(  -{\frac{G_{A}-G_{B}}{RT}}\right)  .
\]
The dashed line means that the stability and the small-scale dynamics of
domains and subunits in the content of A and B conformers can differ from each
other. The $\left[  A\Leftrightarrow B\right]  $ pulsations are accompanied by
reversible sorption-desorption of $\left(  20-50\right)  \,H_{2}O$ molecules
from the cavities.

Structural domains are space-separated formations with a mass of
$(1-2)\cdot10^{3}$ D. Protein subunits $(MM\ge2\cdot10^{3}D)$, as a rule,
consist of 2 or more domains. The domains can consist only of $\alpha$ or only
of $\beta$-structure or have no like secondary structure at all (Schulz,
Schirmer, 1979).

\begin{center}
The shift of $A\Leftrightarrow B$ equilibrium of central cavity of oligomeric
proteins determines their cooperative properties during consecutive ligand
binding in the active sites. Signal transmission from the active sites to the
remote regions of macromolecules is also dependent on $\left(
A\Leftrightarrow B\right)  \,\,$equilibrium.
\begin{center}
\includegraphics[
height=2.3488in,
width=4.9528in
]%
{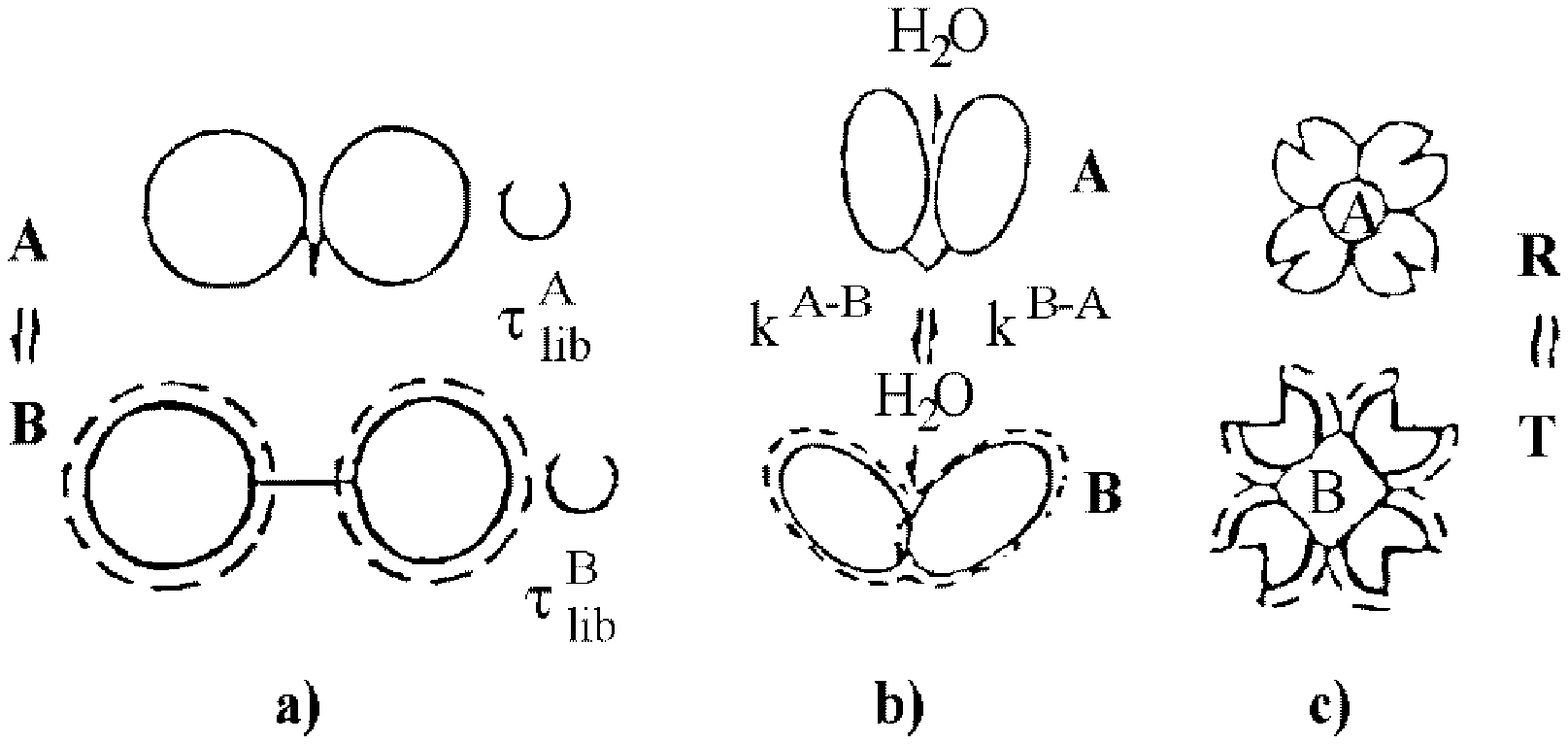}%
\end{center}
\medskip
\end{center}

\begin{quotation}
\textbf{Fig.~1. }Examples of large-scale (LS) protein dynamics:
$A\Leftrightarrow B$ pulsations and librations with correlation times
$(\tau_{\text{lb}}^{B}<\tau_{\text{lb}}^{A})\;($K\"{a}iv\"{a}r\"{a}inen, 1985, 1989):

a) mobility of domains connected by flexible hinge or contact region, like in
the light chains of immunoglobulins;

b) mobility of domains that form the active sites of proteins, like in
hexokinase, papain, pepsin, lysozyme etc. due to flexibility of contacts;

c) mobility of subunits forming the oligomeric proteins like hemoglobin.
Besides transitions of the active sites of each subunit, the $\left(
A\Leftrightarrow B\right)  $ pulsations with frequencies of $(10^{4}-10^{6})$
$s^{-1}$ are pertinent to the common central cavity.

b) \textit{librations }represent the relative rotational - translational
motions of domains and subunits in composition of A and B conformers with
correlation times $\tau_{M}\simeq(1-5)\cdot10^{-8}$s.
\end{quotation}

\medskip

\textbf{LS - \thinspace librations of domains are accompanied by
''flickering'' of water cluster in the open cavity between domains or
subunits. The process of water cluster ''flickering'', i.e. [dissociation
}$\rightleftharpoons$\textbf{association]\ is close to the reversible
first-order phase transition, when:}%

\[
\Delta G_{H_{2}O}=\Delta H_{H_{2}O}-T\Delta S_{H_{2}O}\approx0
\]

\textbf{Such type of transitions in water-macromolecular systems could be
responsible for so called ''enthalpy-entropy compensation effects'' (Lumry and
Biltonen, 1969).}

\textbf{The ''flickering clusters'' means excitation of [}$lb/tr]\,\,$%
\textbf{conversions between librational and translational primary water
effectons, accompanied by [association/dissociation] of coherent water cluster
(see difference in dimensions of lb and tr effectons on Fig. }$18a,b\,\,$%
\textbf{of [1]}$)$\textbf{.}

\textbf{The water cluster (primary lb effecton) association and dissociation
in protein cavities in terms of mesoscopic model represent the }%
$(ac)$\textbf{\ - convertons or }$\left(  \mathit{bc}\right)  $\textbf{\ -
convertons. These excitations stimulate the LS- librations of domains in
composition of B-conformer. The frequencies of (ac) and (bc) convertons, has
the order of about }$10^{8}c^{-1}$\textbf{. This value coincides well with
experimental characteristic times for protein domains librations.}

\textbf{The (ac) and (bc) convertons represent transitions between similar
states of primary librational and translational effectons: }$[a_{lb}%
\rightleftharpoons a_{tr}]$\textbf{\ and }$[b_{lb}\rightleftharpoons b_{tr}%
]$\textbf{\ (see Introduction to[1, 2]).}

\textbf{For the other hand, the Macroconvertons, representing simultaneous
excitation of }$\left(  ac+bc\right)  $\textbf{\ convertons, are responsible
for }$\left[  B\rightleftharpoons A\right]  \,\,$\textbf{large-scale
pulsations of proteins.}

\textbf{\medskip}

The librational mobility of domains and subunits is revealed by the fact that
the experimental value of $\tau_{M}$ is less than the theoretical one
($\tau_{M}^{t}$) calculated on the Stokes-Einstein formula:
\[
\tau_{M}^{t}=(V/k)\cdot\eta/T
\]

This formula is based on the assumption that the whole protein can be
approximated by a rigid sphere. It means, that the large-scale dynamics can be
characterized by the ''flexibility factor'', in the absence of aggregation
equal to ratio:%

\[
fl=(\tau_{M}/\tau_{M}^{t})\leq1
\]
Antonchenko (1986) has demonstrated, using the Monte-Carlo method for
simulations, that the \textbf{disjoining pressure} of a liquid in the pores
onto the walls changes periodically depending on the distance (\textbf{L})
between the limiting surfaces. If the water molecules are approximated by
rigid globes, then the maxima of the wedging pressure lie on the values of
distance \textbf{L}$:9.8;7;$ and $3.3\AA$. It points, that small changes in
the geometry of cavities can lead to significant changes in their
$A\Leftrightarrow B$ equilibrium constant $(K_{A\Leftrightarrow B})$.

\smallskip

According to our model the large-scale transition of the protein cavity from
the ''open'' B-state to the ''closed'' A-state consists of the following stages:

1.~Small reorientation (libration) of domains or subunits, which form an
''open'' cavity (B-state). This process is induced by $\left(  ac\right)  $ or
$\left(  bc\right)  $ convertons of water librational effecton, localized in
cavity (flickering of water cluster);

2.~Cavitational fluctuation of water cluster, containing ($20-50)\,\,\,H_{2}O
$ molecules and the destabilization of the B-state of cavity as a result of
[lb$\rightleftharpoons$tr] \textbf{macroconverton} excitation;

3.~Collapsing of a cavity during the time about $10^{-10}\,s$, dependent on
previous stage and concomitant rapid structural change in the hinge region of
interdomain and intersubunit contacts: $\left[  B\rightarrow A\right]  $ transition.

\smallskip

\textit{The }$b\rightarrow a$ \textit{transition of one of the protein
cavities }can be followed by similar or the opposite $A\rightarrow B$
transition of the other cavity in the macromolecule.

It should be noted that the collapsing time of a cavitation bubble with the
radius: $r\simeq(10-15)\AA$ in \textbf{bulk} water and collapsing time of
interdomain cavity are of the same order: $\Delta t~\symbol{126}10^{-10}s$
under normal conditions$\,($Shutilov, 1980).

If configurational changes of macromolecules at $B\rightarrow A$
and\thinspace$A\rightarrow B$ transitions are sufficiently quick and occur as
a jumps of the effective volume, they accompanied by \textit{appearance of the
shock acoustic waves in the bulk medium.}

When the cavitational fluctuation of water in the ''open'' cavity does not
occur, then $(b\rightarrow a)$ or $(B\rightarrow A)$ transitions are slower
processes, determined by continuous diffusion of domains and subunits. This
happens when $[lb\rightarrow tr]$ macroconvertons are not excited.

In their review, Karplus and McCammon (1986) analyzed data on
alcoholdehydrogenase, myoglobin and ribonuclease, which have been obtained
using \textit{molecular dynamics }approach. It has been shown that large-scale
reorientation of domains occur together with their deformation and motions of
$\alpha$ and $\beta$ structures.

It has been shown also (Karplus and McCammon, 1986) that activation free
energies, necessary for $\left[  A\Leftrightarrow B\right]  $ transitions and
the reorganization of hinge region between domains, do not exceed (3-4)
kcal/mole. Such low values were obtained for proteins with even rather dense
interdomain region, as seen from X-ray data. The authors explain such low
values of activation energy by the fact that the displacement of atoms,
necessary for such transition, does not exceed 0.5 \AA, i.e. they are
comparable with the usual amplitudes of atomic oscillation at temperatures
$20-30^{0}$C. It means that they occur very quickly within times of
$10^{-12}\,s$, i.e. much less than the times of $\left[  A\rightarrow
B\right]  $ or $\left[  B\rightarrow A\right]  $ domain displacements
$(10^{-9}-10^{-10}s)$. Therefore, the high frequency small-scale dynamics of
hinge is responsible for the quick adaptation of hinge geometry to the
changing distance between the domains and for decreasing the total activation
energy of $\left[  A\Leftrightarrow B\right]  $ pulsations of proteins,.

Recent calculations by means of molecular dynamics reveal that the
oscillations in proteins are harmonic at the low temperature (T$<220K)$ only.
\textbf{At the physiological temperatures the oscillations are strongly
unharmonic, collective, global and their amplitude increases with hydration
(Steinback et al., 1996). }{\large Water is a ''catalyzer'' of protein
unharmonic dynamics.}

\textbf{It is obvious, that both small-scale (SS) and large-scale (LS)
dynamics, introduced in our model, are necessary for protein function. To
characterize quantitatively the LS dynamics of proteins, we proposed the
unified Stokes-Einstein and Eyring-Polany equation.}

\medskip

\begin{center}
\smallskip

{\large 2. Description of large-scale dynamics of proteins, based on
generalized Stokes-Einstein and Eyring-Polany equation}

\smallskip
\end{center}

In the case of the \textbf{continuous }Brownian diffusion of a particle, the
rate constant of diffusion is determined by the Stokes-Einstein law:%

\begin{equation}
k={\frac{1}{\tau}}=\frac{k_{B}T}{V\eta}\tag{1}%
\end{equation}
where: $\tau$ is correlation time, i.e. the time, necessary for rotation of a
particle by the mean angle determined as $\bar{\varphi}\approx0.5$ of the turn
or the characteristic time for the translational movement of a particle with
the radius (a) on the distance $(\bar{\Delta}_{x})^{1/2}\simeq0.6a\;($%
Einstein, 1965);

$V=4\pi a^{3}/3\;$ is the volume of the spherical particle; $k_{B}$ is the
Boltzmann constant, T and $\eta$ are the absolute temperature and bulk
viscosity of the solvent.

On the other hand, the rate constant of$\;\left[  A\rightarrow B\right]  $
reaction for a molecule in gas phase, which is related to passing through the
activation barrier $G^{A\rightarrow B}$, is described with the Eyring-Polany equation:%

\begin{equation}
k^{A\rightarrow B}={\frac{kT_{B}}{h}}\exp\left(  -{\frac{G^{A\rightarrow B}%
}{RT}}\right) \tag{2}%
\end{equation}
To describe the large-scale dynamics of macromolecules in solution related to
fluctuations of domains and subunits (librations and pulsations), an equation
is needed which takes into account the diffusion and activation processes simultaneously.

The rate constant for the rotational- translational diffusion of the particle
$(k_{c})$ forming a macromolecule (continuous LS-dynamics) is expressed with
the generalized Stokes-Einstein and Eyring-Polany equation (K\"aiv\"ar\"ainen
and Goryunov, 1987):%

\begin{equation}
K_{c}={\frac{k_{B}T}{\eta V}}\exp\left(  -{\frac{G_{st}}{RT}}\right)
=\tau_{c}^{-1}\tag{3}%
\end{equation}

where: V is the effective volume of domain or subunit, which are capable to
the Brownian mobility independently from the rest part of the macromolecule,
with the probability:%

\begin{equation}
P_{lb}=\exp\left(  -{\frac{G_{st}}{RT}}\right)  ,\tag{4}%
\end{equation}
where: G$_{st}$ is the activation energy of structural change in the contact
(hinge) region of a macromolecule, necessary for independent mobility of
domain or subunit; $\tau_{c}$ is the effective correlation time for the
continuous diffusion of this relatively independent particle.

\smallskip

\textbf{The effective volume V can be changed under the influence of
temperature, perturbants and ligands.}

\textbf{The generalized Stokes-Einstein and Eyring-Polany equation (3) is
applicable also to describing the diffusion of the whole (integer) particle,
dependent on the surrounding medium fluctuations with activation energy
}$(G_{a})$\textbf{. The ligand diffusion in the active site cavity of proteins
is such a type of processes.}

\smallskip

\textbf{To describe noncontinuous process,} the formula for rate constant
$(k_{\text{jump}})$ of the jump-like translations of particle, related to
emergency of cavitational fluctuations (holes) near the particle was proposed
(K\"{a}iv\"{a}r\"{a}inen and Goryunov, 1987):%

\begin{equation}
k_{\text{jump}}={\frac{1}{\tau_{\text{jump}}^{\min}}}\exp\left(  -{\frac
{W}{RT}}\right)  ={\frac{1}{\tau_{\text{jump}}}},\tag{5}%
\end{equation}

where:%

\begin{equation}
W=\sigma S+n_{s}(\mu_{\text{out}}-\mu_{\text{in}})\tag{6}%
\end{equation}
is the work of \textit{cavitation fluctuation }with the cavity surface S, at
which $n_{s}$ molecules of the solvent (water) change its effective chemical
potential from $\mu_{\text{in}}$ to $\mu_{\text{out}}$.

The dimensions of cavity fluctuation near particle must be comparable to
corresponding particles.

In a homogeneous phase (i.e. pure water) under equilibrium conditions we have:
$\mu_{\text{in}}=\mu_{\text{out}}$. With an increase of particle sizes,
surface of cavitational fluctuation (S) and its work (W), the corresponding
probability of cavitation fluctuations:%

\[
P_{\text{jump}}=\exp(-W/RT)
\]
will fall.

The notion of the surface energy ($\sigma$) retains its meaning even at very
small ''holes'' because of its molecular nature (see Section 11.4 of [1] and [4]).

$\tau_{\text{jump}}^{\min}$ in eq. (5) is the minimal possible jump-time of a
particle with mass (m) over the distance $\lambda$ with the mean velocity:%

\begin{equation}
v_{\max}=(2kT/m)^{1/2}\tag{7}%
\end{equation}
Hence, we derive for the maximal jump-rate at W=0:%

\begin{equation}
k_{\text{jump}}^{\max}={\frac{1}{\tau_{\text{jump}}^{\min}}}={\frac{V_{\max}%
}{\lambda}}={\frac{1}{\lambda}}\left(  {\frac{2kT}{m}}\right)  ^{1/2}\tag{8}%
\end{equation}
In the case of hinged domains, forming macromolecules their relative
$A\rightleftharpoons B$ displacements (pulsations) are related not only to
possible holes forming in the interdomain (intersubunit) cavities or near
their outer surfaces, but to the structural change of hinge regions as well.

If the activation energy of necessary structure changes is equal to
G$_{st}^{A\rightleftharpoons B}$, then eq. (5), with regard for (8), is
transformed into%

\begin{equation}
k_{\text{jump}}^{A\Leftrightarrow B}={\frac{1}{\lambda}}\left(  {\frac{2kT}%
{m}}\right)  ^{1/2}\exp\left(  -{\frac{W_{A,B}+G_{st}^{A\Leftrightarrow B}%
}{RT}}\right) \tag{9}%
\end{equation}

where: W$_{A,B}$ is the work required for cavitational fluctuations of water;
this work can be different in two directions: $(W_{B})$ is necessary for
nonmonotonic $B\rightarrow A$ transition; $(W_{A})$ is necessary for jump-way
$A\rightarrow B$ transition.

Under certain conditions $A\rightleftharpoons B$ transitions between protein
conformers (LS- pulsations) can be realized owing to the \textit{jump}%
-\textit{way and continuous }types of relative diffusion of domains or
subunits as two stage reaction. In this case, the resulting rate constant of
the process will be expressed through (9) and (3) as:%

\begin{equation}%
\begin{array}
[c]{l}%
k_{\text{res}}^{A\Leftrightarrow B}=k_{\text{jump}}^{A\Leftrightarrow B}%
=k_{c}^{A\Leftrightarrow B}={\frac{1}{\lambda}}\left(  {\frac{2kT}{m}}\right)
^{1/2}\exp\left(  -{\frac{W_{A,B}+G_{st}^{A\Leftrightarrow B}}{RT}}\right)
+\\
\\
+{\frac{kT}{\lambda V}}\exp\left(  -{\frac{G_{st}^{A\Leftrightarrow B}}{RT}%
}\right)
\end{array}
\tag{10}%
\end{equation}

The interaction between two domains in \textit{A-conformer }can be described
using microscopic Hamaker - de Bour theory. One of the contributions into
$G_{st}^{A\Leftrightarrow B}$ is the energy of dispersion interactions between
domains of the \textit{radius (a)} (K\"{a}iv\"{a}r\"{a}inen, 1989b):%

\begin{equation}
\lbrack G_{st}\sim U_{H}\approx-A^{*}a/12H]_{A,B}\tag{11}%
\end{equation}
where%

\begin{equation}
A^{*}\approx(A_{s}^{1/2}-A_{c}^{1/2})^{2}\approx{\frac{3}{2}}\pi h\nu_{0}%
^{s}\left[  \alpha_{s}N_{s}-\alpha_{c}N_{c}\right]  ^{2}\tag{12}%
\end{equation}
is the complex Hamaker constant; H is the slit thickness between domains in
\textit{A-state}; $\;A_{c}$ and $A_{s}$ are simple Hamaker constants,
characterizing the properties of water in the \textit{A-state} of the cavity
and in the bulk solvent, correspondingly. They depend on the concentration of
water molecules $(N_{c}\simeq N_{s})$ and their polarizability $(\alpha
_{c}\neq\alpha_{s})$:%

\[
A_{c}={\frac32}\pi h\nu_{0}^{c}\alpha_{c}^{2}N_{c}^{2}\;\;\text{
and\ \ \ }A_{s}={\ \frac32}\pi h\nu_{0}^{s}\alpha_{s}^{2}N_{s}^{2};
\]
where: $h\nu_{0}^{c}\approx h\nu_{0}^{s}$ are the ionization potentials of
$H_{2}O$ molecules in a cavity and in a free solvent.

\textit{In the ''closed'' A-state of a cavity the water layer between domains
has a more compact packing as compared with the ice-like structure of a water
cluster in the B-state of a cavity, or with a free solvent. As far
$H_{A}<H_{B}$ the dispersion interaction (11) between domains in A- state of
cavity is stronger, than that in B-state: $U_{H}^{A}>U_{H}^{B}$.}

Disjoining pressure of water in the cavities
\begin{equation}
\Pi=-A^{*}/6\pi H^{3}\tag{12a}%
\end{equation}
decreases with the increase of the complex Hamaker constant $(A^{*})$ that
corresponds to the increase of the attraction energy $(U_{H})$ between domains.

Cooperative properties of clusters in open (B)-states of the cavities are more
pronounced as compared to that in bulk water. That results in the greater
changes of $\alpha_{c}N_{c}$ than that of $\alpha_{s}N_{s}$ induced by
temperature. The elevation of temperature decreasing the dimensions of
interdomain water clusters leads to the strengthening of interdomain
interaction, while the lowering temperature leads to opposite effect.

We can judge about the changes of $\,\,\alpha_{s}$N$_{s}\,\,$ in the
experiment on measuring the solvent refraction index, as far from our theory
of refraction index (eq. 8.14 of [1] or paper [3]):%

\begin{equation}
(n_{s}^{2}-1)/n_{s}^{2}={\frac{4}{3}}\pi\alpha_{s}N_{s}\text{ \ \ \ or:\ \ \ }%
\alpha_{s}N_{s}={\frac{3}{4}}\pi\cdot{\frac{n_{s}^{2}-1}{n_{s}^{2}}}\tag{12b}%
\end{equation}

In the \textbf{closed cavities }the effect of temperature on water properties
is lower as compared to that in bulk water. \textbf{It follows that
thermoinduced nonmonotonic transition in the solvent refraction index must be
accompanied by in-phase nonmonotonic changes of the }$\left[
\mathbf{A\Leftrightarrow B}\right]  $\textbf{\ equilibrium constant
$(K_{A\Leftrightarrow B})$.} As far (A) and (B) conformers usually have
different stability and flexibility, the changes of $K_{A\Leftrightarrow B}$
will be manifested in the changes of protein large-scale and small-scale
dynamics. It has been shown before that viscosity itself has nonmonotonic
temperature dependence due to the nonmonotonic dependence of $n^{2}%
(t)\;(eq.11.44,11.45$ and 11.48 of [1] or paper [4]).

\textbf{Thus, thermoinduced non-denaturational transitions of macromolecules
and supramolecular systems located in the aqueous environment are caused by
nonmonotonic changes in solvent properties, including its refraction index.}

The influence of $D_{2}O$ and other perturbants on protein dynamics is
explained in a similar way. The effect of deuterium oxide ($D_{2}O)$ is a
result of substitution of $H_{2}O$ from protein cavities and corresponding
change of complex Hamaker constant (12).

Generalized equation (3) is applicable not only for evaluating the frequency
of macromolecules transition between A and B conformers but also for the
frequency of the dumped librations of domains and subunits within A and B
conformers. Judging by various data (K\"{a}iv\"{a}r\"{a}inen, $1985,1989b)$,
the interval of $A\rightleftharpoons B$ pulsation frequency is:%

\begin{equation}
\nu_{A\Leftrightarrow B}={\frac{1}{t_{A}+t_{B}}}\approx k^{A\rightarrow
B}=(10^{4}-10^{7})\text{ }s^{-1}\tag{13}%
\end{equation}
where: $t_{A}$ and $t_{B}$ are the lifetimes of A and B conformers.

The corresponding interval of the total activation energy of the jump-way
$A\Leftrightarrow B$ pulsations can be evaluated from the eq. (9). We assume
for this end that the pre-exponential multiplier is about $10^{10}s^{-1}$ as a
frequency of cavitational fluctuations in water with the radius \symbol{126}%
(10-15) $\overset{o}{A}$.

Taking a logarithm of (9) we derive:%

\begin{equation}
G_{\text{res}}^{A\Leftrightarrow B}=(W_{A,B}+G_{st}^{A\Leftrightarrow
B})\approx RT(\ln10^{10}-\ln\nu_{A\Leftrightarrow B})\tag{14}%
\end{equation}
At physiological temperatures the following region of energy corresponds to
the frequency range of pulsations (13)%

\begin{equation}
G_{\text{res}}^{A\Leftrightarrow B}\approx(4-8)\text{ kcal/mole}\tag{15}%
\end{equation}
Such a region of energies is pertinent to a wide range of biochemical processes.

The quick jump-way pulsations of macromolecules can cause acoustic shock-
waves in the solvent and its structure destabilization. Concomitant increase
in water activity leads to distant interaction between different proteins as
well as proteins and cells. Such solvent-mediated phenomena were discovered
and studied in our laboratory by set of specially elaborated methods
(K\"aiv\"ar\"ainen, 1985, 1986, 1987; K\"aiv\"ar\"ainen et al., 1990,
K\"aiv\"ar\"ainen et al., 1993).

When the $\left[  A\Leftrightarrow B\right]  $ transitions in proteins are
related to \textit{continuous }diffusion only, then the $G_{st}$ values
calculated using eq.(3) for the same frequency interval $(10^{4}-10^{7})$ $s$,
is about (3 - 7) kcal/mole.

\smallskip

The \textit{Kramers equation }(1940), which has earlier been widely used for
describing diffusion processes, has the form:%

\begin{equation}
k={\frac{A}{\eta}}\exp\left(  -{\frac{H^{*}}{RT}}\right) \tag{16}%
\end{equation}
where A is a constant and $\eta$ - solvent viscosity.

\smallskip

\textbf{The pre-exponential factor in our generalized equation (3) contains
not only the viscosity, but also the temperature and the effective volume of a
particle. It was shown in our experiments that eq. (3) describes the dynamic
processes, which occur in macromolecules, solutions much better than the
Kramer's equation (16).}\bigskip

\begin{center}
\medskip{\large 3. Dynamic model of protein-ligand complexes formation}
\end{center}

\smallskip

\textbf{According to our model of specific complexes formation the following
order of events is assumed (Fig. 2):}

\textbf{1.~Ligand (L) collides with the active site (AS), formed usually by
two domains, in its open (b) state: the structure of water cluster in AS is
being perturbed and water is forced out of AS cavity totally or partially;}

\textbf{2.~Transition of AS from the open (b) to the closed (a) state occurs
due to strong shift of }$[a\Leftrightarrow b]$\textbf{\ equilibrium to the
left, i.e. to the AS domains large scale dynamics;}

\textbf{3.~A process of dynamic adaptation of complex [L+AS] begins,
accompanied by the directed ligand diffusion in AS cavity due to its domains
small-scale dynamics and deformation of their tertiary structure;}

\textbf{4.~If the protein is oligomeric with few AS, then the above events
cause changes in the geometry of the central cavity between subunits in the
open state leading to the destabilization of the large central water cluster
and the shift of the }$A\rightleftharpoons B,$\textbf{\ corresponding to
}$R\rightleftharpoons T$\textbf{\ equilibrium of quaternary structure
leftward. Water is partially forced out from central cavity.}

\textbf{Due to the feedback mechanism this shift can influence the }$\left[
a\Leftrightarrow b\right]  $\textbf{\ equilibrium of the remaining free AS and
promotes its reaction with the next ligand. Every new ligand stimulates this
process, promoting the positive cooperativity. The negative cooperativity also
could be resulted from the interaction between central cavity and active sites;}

\textbf{5.~The terminal }$\left[  protein-ligand\right]  $\textbf{\ complex is
formed as a consequence of the relaxation process, representing deformation of
domains and subunits tertiary structure. This stage could be much slower than
the initial ones [1-3]. As a result of it, the stability of the complex grows up.}

\smallskip

\textbf{Dissociation of specific complex }is a set of reverse processes to
that described above which starts from the $\left[  a^{*}\rightarrow b\right]
$ fluctuation of the AS cavity.

\smallskip

\textbf{In multidomain proteins like antibodies, which consist of 12 domains,
and in oligomeric proteins, the cooperative properties of }$H_{2}%
O$\textbf{\ clusters in the cavities can determine the mechanism of signal
transmission from AS to the remote effector regions and allosteric protein properties.}

The stability of a librational water effecton as coherent cluster strongly
depends on its sizes and geometry. This means that very small deformations of
protein cavity, which violate the [cavity-cluster] complementary condition,
induce a cooperative shift of $\left[  A\Leftrightarrow B\right]  $
equilibrium leftward. The clusterphilic interaction, introduced by us (see
section 13.3 of [1] or paper [4]) turns to hydrophobic one due to $[lb/tr]$ conversion.

\textbf{This process can be developed step by step}. For example, the
reorientation of variable domains, which form the antibodies active site (AS)
after reaction with the antigen determinant or hapten deforms the next cavity
between pairs of variable and constant domains forming F$_{ab}$ subunits
(Fig.2). The leftward shift of $\left[  A\Leftrightarrow B\right]  $
equilibrium of this cavity, in turn, changes the geometry of the big central
cavity between $F_{ab}$ and $F_{c}$ subunits, perturbing the structure of the
latter.\textbf{\ Therefore, the signal transmission from the AS to the
effector sites of $F_{c}$ subunits occurs due to the balance shift between
clusterphilic and hydrophobic interactions. }This signal is responsible for
complement- binding sites activation and triggering the receptors function on
the lymphocyte membranes, in accordance to our model.

The leftward shift of $\left[  A\Leftrightarrow B\right]  $ equilibrium in a
number of cavities in the elongated multidomain proteins can lead to the
significant decrease of their linear size and dehydration. The mechanism of
muscular contraction is probably based on such phenomena and clusterphilic
interactions (see next section).

For such a nonlinear system the energy is necessary for reorientation of the
first couple of domains only. The process then goes on spontaneously with
decreasing the averaged protein chemical potential.

The chemical potential of the A- conformer is usually lower than that of
B-\thinspace\thinspace conformer $(\bar{G}_{A}<\bar{G}_{B})$ and the
relaxation of protein is accompanied by the leftward $A\Leftrightarrow B$
equilibrium shift of cavities.

It is predictable, that hydration of proteins will decrease, when
clusterphilic [water-cavity] interaction turns to hydrophobic one.

\begin{center}%
\begin{center}
\includegraphics[
height=2.9049in,
width=4.5809in
]%
{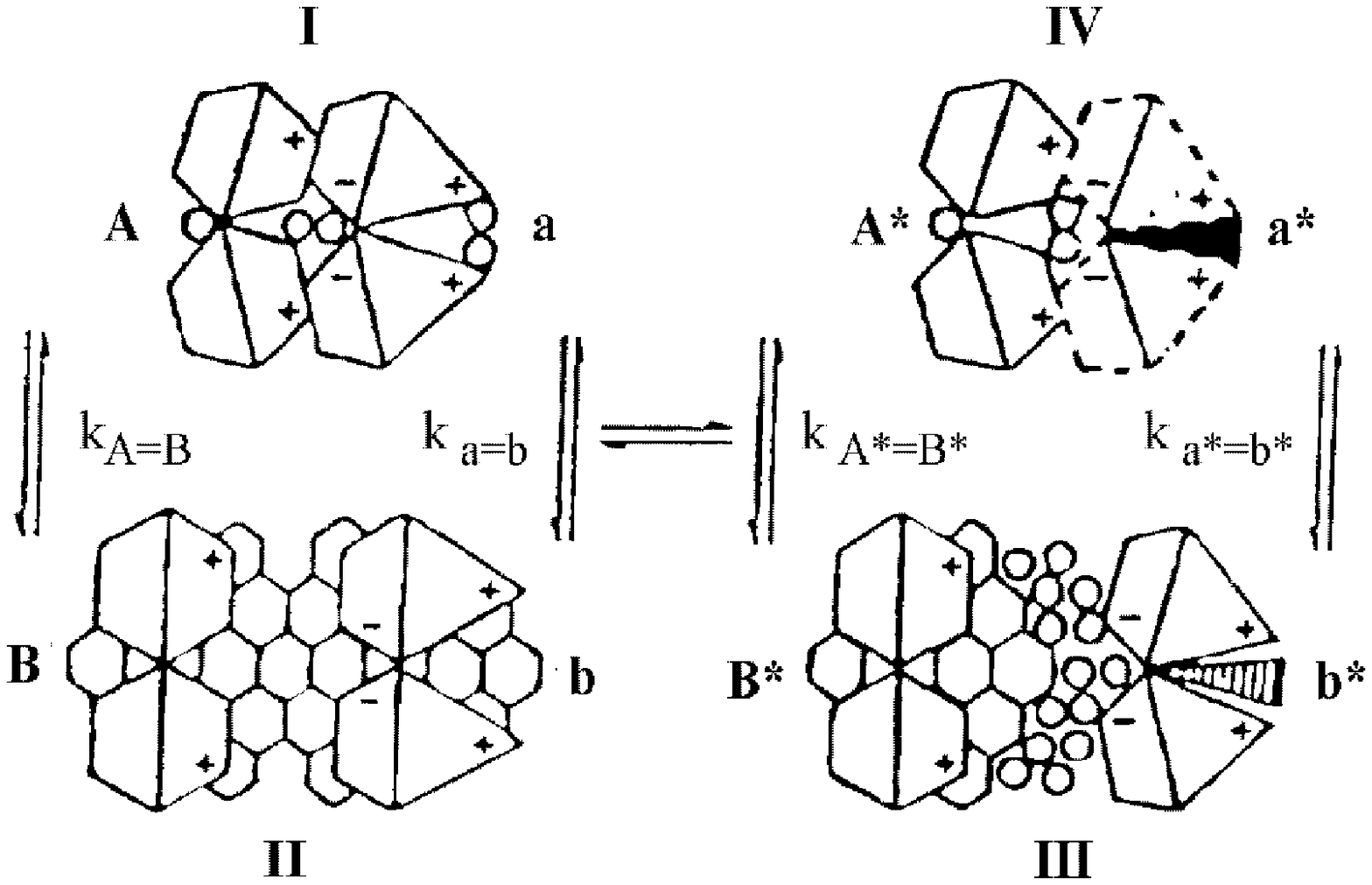}%
\end{center}
\medskip
\end{center}

\begin{quotation}
\textbf{Fig.~2. }The schematic picture of the protein association (Fab
subunits of antibody with a ligand), which is accompanied by the
destabilization of water clusters in cavities, according to the dynamic model
(K\"{a}iv\"{a}r\"{a}inen, 1985). The dotted line denotes the perturbation of
the tertiary structure of the domains forming the active site. Antibodies of
IgG type contain usually two such Fab subunit and one Fc subunit, conjugated
with 2Fab by flexible hinge, forming the general Y-like structure.
\end{quotation}

\medskip

Our dynamic model of protein behavior and signal transmission, described
above, is an alternative to solitonic mechanism of non dissipative signal
transmission in proteins and in other biosystems proposed by Davidov (1973).
Propagation of solitonic wave is a well known nonlinear process in the ordered
\textbf{homogeneous} mediums. The solitons can originate, when the nonlinear
effects are compensated by the wave dispersion effects. Dispersion is
reflected in fact that the longer waves spreads in medium with higher velocity
than the shorter ones.

\textbf{\ However, biosystems of nonregular, fluctuating structure are not the
mediums, good for solitons emergency and propagation.}

\textbf{\ Our dynamic model takes into account the real multidomain and
multiglobular structure of a proteins and properties of their hydration shell
fractions. In contrast to Davidov's solitonic model, the dissipation processes
like reversible ''melting'' of water clusters, accompanied by large-scale
dynamics of proteins, are the necessary stages of our [hydrophobic
}$\rightleftharpoons\,$\textbf{clusterphilic] mechanism of signal transmission
in biosystems.}

\medskip

\textbf{The evolution of the ideas of the protein-ligand complex formation
proceeded in the following sequence:}

\textbf{1.~''Key-lock'' or the rigid conformity between the geometry of an
active site and that of a ligand (Fisher, 1894);}

\textbf{2.~''Hand-glove'' or the so-called principle of induced conformity
(Koshland, 1962);}

\textbf{3.~At the current stage of complex-formation process understanding,
the crucial role of protein dynamics gets clearer. Our model allows us to put
forward the ''Principle of Stabilized Conformity (PSC)'' instead that of
''induced conformity'' in protein-ligand specific reaction.}

\smallskip

{\large Principle of Stabilized Conformity (PSC) means that the geometry of
the active site (AS), optimal from energetic and stereochemical conditions, is
already existing BEFORE reaction with ligand. The optimal geometry of AS is to
be the only one selected among the number of others and stabilized by ligand,
but not induced ''de nova''.}

\smallskip

\textbf{For example, the }$[a\Leftrightarrow b]$\textbf{\ large-scale
pulsations of the active sites due to domain fluctuations and stabilization of
the closed (a) state by ligand are necessary for the initial stages of
reaction. Such active site pulsations decreases the total activation energy
necessary for the terminal complex formation as multistage process.}

\textbf{\medskip}

\begin{center}
{\large 4. The life-time of quasiparticles and frequencies of their excitation}

\textbf{\smallskip}
\end{center}

\textbf{The set of formula, describing the dynamic properties of
quasiparticles, introduced in mesoscopic theory was presented at Chapter 4 of
book [1] and paper [2]:}

\textbf{\medskip}

\textit{The frequency of c- Macrotransitons or Macroconvertons excitation,
representing} \textit{[dissociation/association] of primary librational
effectons }- ''\textit{flickering clusters}'' \textit{as a result of
interconversions between primary [lb] and [tr] effectons is: }%

\begin{equation}
F_{cM}={\frac{1}{\tau_{Mc}}}\cdot P_{Mc}/Z\tag{17}%
\end{equation}

where: $P_{Mc}=P_{ac}\cdot P_{bc}\;$ is a probability of macroconverton excitation;

$Z\;$ is a total partition function (see eq.4.2 of [1, 2]);

the life-time of macroconverton is:%

\begin{equation}
\tau_{Mc}=(\tau_{ac}\cdot\tau_{bc})^{1/2}\tag{18}%
\end{equation}

The cycle-period of (ac) and (bc) convertons are determined by the sum of
life-times of intermediate states of primary translational and librational effectons:%

\begin{equation}%
\begin{array}
[c]{c}%
\tau_{ac}=(\tau_{a})_{tr}+(\tau_{a})_{lb};\\
\tau_{bc}=(\tau_{b})_{tr}+(\tau_{b})_{lb};
\end{array}
\tag{19}%
\end{equation}

The\textit{\ }life-times of primary and secondary effectons (lb and tr) in
\textit{a}- and \textit{b}-states are the reciprocal values of corresponding
state frequencies:%

\begin{equation}
\lbrack\tau_{a}=1/\nu_{a};\text{ \thinspace\thinspace}\tau_{\overline{a}%
}=1/\nu_{\overline{a}}]_{tr,lb};\text{ \qquad[}\tau_{b}=1/\nu_{b};\text{
\thinspace\thinspace}\tau\overline{_{b}}=1/\nu_{\overline{b}}]_{tr,lb}\tag{20}%
\end{equation}

[$(\nu_{a})$ and $(\nu_{b})]_{tr,lb}\;$ correspond to eqs. 4.8 and 4.9 of [1, 2];

[$(\nu_{\overline{a}})$ and $(\nu_{\overline{b}})]_{tr,lb}\;$\text{ }could be
calculated using eqs.4.16; 4.17 [1, 2].

\textbf{The frequency of }$\mathbf{(ac)}$\textbf{\ and }$\mathbf{(bc}%
$\textbf{) convertons excitation [lb/tr]:}%

\begin{equation}
F_{ac}={\frac{1}{\tau_{ac}}}\cdot P_{ac}/Z\tag{21}%
\end{equation}%

\begin{equation}
F_{bc}={\frac{1}{\tau_{bc}}}\cdot P_{bc}/Z\tag{22}%
\end{equation}

where: $P_{ac}$ and $P_{bc}$ are probabilities of corresponding convertons
excitations (see eq.4.29a of [1, 2]).

\smallskip

\textbf{The frequency of Supereffectons and Superdeformons (biggest
fluctuations) excitation is:}\textit{\ }%

\begin{equation}
F_{SD}={\frac{1}{(\tau_{A^{*}}+\tau_{B^{*}}+\tau_{D^{*}})}}\cdot P_{S}^{D^{*}%
}/Z\tag{23}%
\end{equation}

\textit{It is dependent on cycle-period of Supereffectons: $\tau_{SD}%
=\tau_{A^{*}}+\tau_{B^{*}}+\tau_{D^{*}}$}

\textit{\ and probability of Superdeformon activation (}$P_{S}^{D^{*}}),$
\textit{like the limiting stage of this cycle.}

The averaged life-times of Supereffectons in $A^{*}$ and $B^{*}$ state are
dependent on similar states of translational and librational macroeffectons :%

\begin{equation}
\tau_{A^{*}}=[(\tau_{A})_{tr}\cdot(\tau_{A})_{lb}]=[(\tau_{a}\tau
_{\overline{a}})_{tr}\cdot(\tau_{a}\tau_{\overline{a}})_{lb}]^{1/2}\tag{24}%
\end{equation}

and that in B state:%

\begin{equation}
\tau_{B^{*}}=[(\tau_{B})_{tr}\cdot(\tau_{B})_{lb}]=[(\tau_{b}\tau
_{\overline{b}})_{tr}\cdot(\tau_{b}\tau_{\overline{b}})_{lb}]^{1/2}\tag{25}%
\end{equation}

\textit{The life-time of Superdeformons excitation }is determined by frequency
of beats between A$^{*}$ and B$^{*}$states of Supereffectons as:%

\begin{equation}
\tau_{D^{*}}=1/\left|  (1/\tau_{A^{*}})-(1/\tau_{B^{*}})\right| \tag{26}%
\end{equation}

\textbf{\ The frequency of translational and librational macroeffectons}
$A\rightleftharpoons B$ cycle excitations could be defined in a similar way:%

\begin{equation}
\left[  F_{M}={\frac{1}{(\tau_{A}+\tau_{B}+\tau_{D})}}\cdot P_{M}%
^{D}/Z\right]  _{tr,lb}\tag{27}%
\end{equation}

where:
\begin{equation}
(\tau_{A})_{tr,lb}=[(\tau_{a}\cdot\tau_{\overline{a}})_{tr,lb}]^{1/2}\tag{28}%
\end{equation}

and
\begin{equation}
(\tau_{B})_{tr,lb}=[(\tau_{b}\cdot\tau_{\overline{b}})_{tr,lb}]^{1/2}\tag{29}%
\end{equation}%

\begin{equation}
(\tau_{D})_{tr,lb}=1/\left|  (1/\tau_{A})-(1/\tau_{B})\right|  _{tr,lb}%
\tag{30}%
\end{equation}

\textbf{The frequency of primary translational effectons }%
$(a\rightleftharpoons$\textbf{$b)_{tr}$ transitions could be expressed like:}%

\begin{equation}
F_{tr}={\frac{1/Z}{(\tau_{a}+\tau_{b}+\tau_{t})_{tr}}}\cdot(P_{d}%
)_{tr}\tag{31}%
\end{equation}
where: $(P_{d})_{tr}$ is a probability of primary translational deformons
excitation (eq. 4.25 of [1, 2]);

$[\tau_{a};\tau_{b}]_{tr}$ are the life-times of (a) and (b) states of primary
\textit{translational} effectons (eq. 20).

\smallskip

\textbf{The frequency of primary librational effectons as
($a\rightleftharpoons b)_{lb}$ cycles excitations is}\textit{:}%

\begin{equation}
F_{lb}={\frac{1/Z}{(\tau_{a}+\tau_{b}+\tau_{t})_{lb}}}\cdot(P_{d}%
)_{lb}\tag{32}%
\end{equation}
where: $(P_{d})_{lb}$is a probability of primary librational deformons
excitation; $\tau_{a}$ and $\tau_{b}$ are the life-times of (a) and (b) states
of primary librational effectons defined as (20).

The life-time of primary transitons (tr and lb) as a result of quantum beats
between (a) and (b) states of primary effectons could be introduced as:%

\begin{equation}
\lbrack{\tau_{t}=\left|  {1/\tau_{a}-1/\tau_{b}}\right|  ^{-1}]_{tr,lb}%
}\tag{33}%
\end{equation}

\smallskip

For the case of $(a\Leftrightarrow b)^{1,2,3}$ transitions of primary and
secondary effectons \textit{(tr \ and\ lb),} their life-times in (a) and (b)
states are the reciprocal value of corresponding frequencies: $[\tau_{a}%
=1/\nu_{a}\;$ and $\;\tau_{b}=1/\nu_{b}]_{tr,lb}^{1,2,3}$.\ These parameters
and the resulting ones could be calculated from eqs.(2.27; 2.28 of [1]) for
primary effectons and (2.54; 2.55 of [1]) for secondary ones.

The results of calculations, using eqs. (31, 32) for frequency of excitations
of primary \textit{tr and lb }effectons are plotted on Fig. 3a,b.

The frequencies of Macroconvertons and Superdeformons were calculated using
eqs.(17 and 23).

\begin{center}%
\begin{center}
\includegraphics[
height=3.4091in,
width=4.0906in
]%
{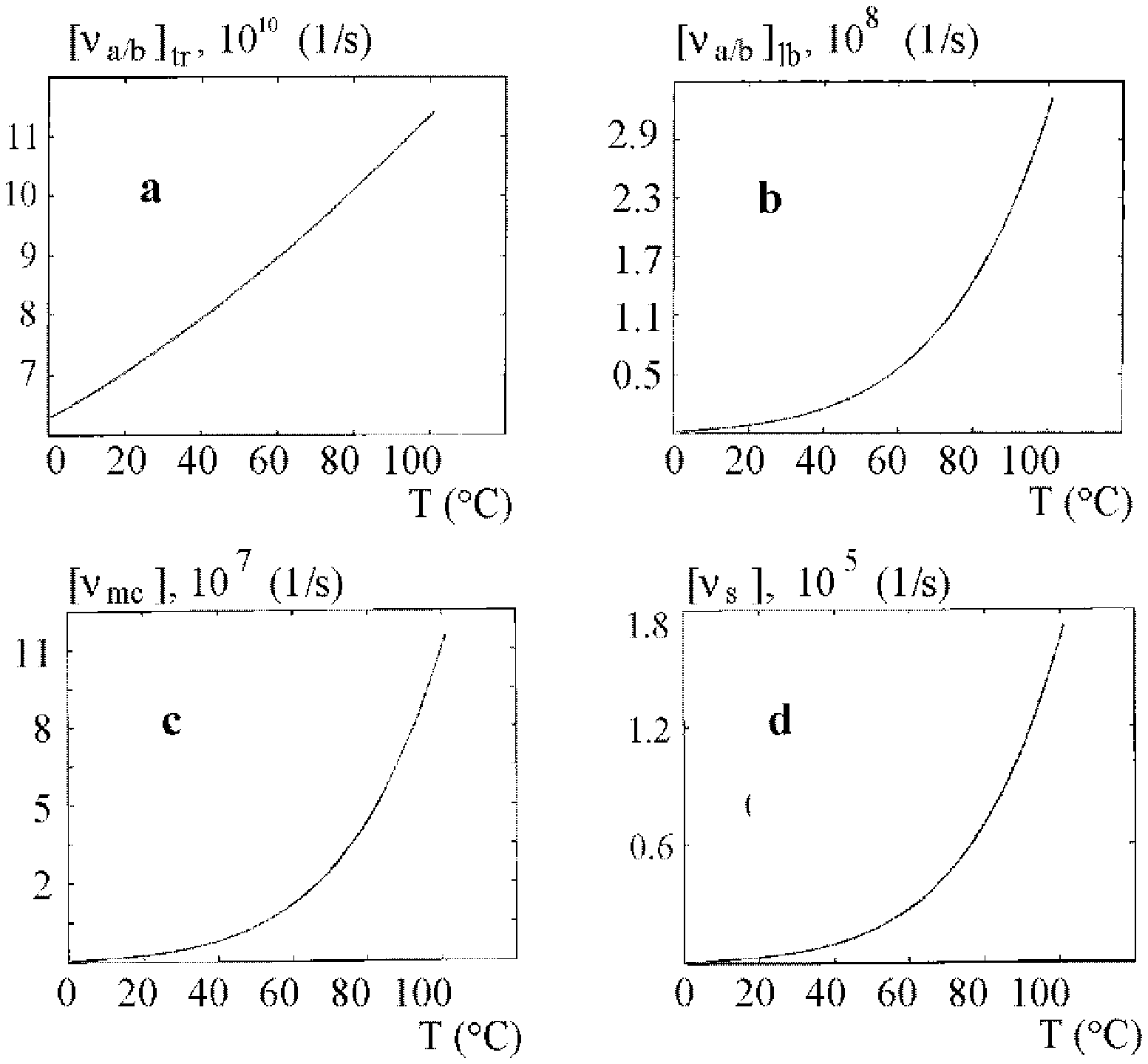}%
\end{center}
\medskip
\end{center}

\begin{quotation}
\textbf{Fig.~3. }(a) - Frequency of primary [tr] effectons excitations,
calculated from eq.(31);

(b) - Frequency of primary [lb] effectons excitations, calculated from eq.(32);

(c) - Frequency of $[lb/tr]$ Macroconvertons (flickering clusters)
excitations, calculated from eq.(17);

(d) - Frequency of Superdeformons excitations, calculated from eq.(28).
\end{quotation}

\medskip

At the temperature interval (0-100)$^{0}C$ the frequencies of translational
and librational macrodeformons (tr and lb) are in the interval of
(1.3-2.8)$\cdot10^{9}s^{-1}$ and (0.2-13)$\cdot10^{6}s^{-1}$ correspondingly.
The frequencies of (ac) and (bc) convertons could be defined also using our
software and formulae, presented at the end of Chapter 4 of [1, 2].

\smallskip

The frequency of primary translational effectons $[a\Leftrightarrow b]_{tr}$
excitations at 20$^{0}C$, calculated from eq.(31) is $\nu\sim7\cdot
10^{10}(1/s)$. It corresponds to electromagnetic wave length in water with
refraction index $(n=1.33)$ of:\
\begin{equation}
\lambda=(cn)/\nu\sim6mm\tag{34}%
\end{equation}
For the other hand, there are a lot of evidence, that irradiation of very
different biological systems with such coherent electromagnetic field exert
great influences on their properties (Grundler and Keilman, 1983).

\smallskip

\textbf{Between the dynamics/function of proteins, membranes, etc. and
dynamics of their aqueous environment the strong interrelation is existing.}

The frequency of macroconvertons, representing big density fluctuation in the
volume of primary librational effecton at 37C is about $10^{7}(1/s)$ $($Fig 3c).

\textbf{The frequency of librational macrodeformons at the same temperature is
about 10}$^{6}$\textbf{\ s}$^{-1},$\textbf{i.e. coincides\ with frequency of
large-scale protein cavities pulsations between open and closed to water
states (see Fig.2). This confirm our hypothesis that the clusterphilic
interaction is responsible for stabilization of the proteins cavities open
state and that transition from the open state to the closed one is induced by
coherent water cluster dissociation.}

\textbf{The frequency of Superdeformons excitation (Fig.3d) is much lower:}%

\begin{equation}
\nu_{s}\sim(10^{4}-10^{5})\text{ }s^{-1}\tag{35}%
\end{equation}
\textbf{Superdeformons are responsible for cavitational fluctuations in
liquids and origination of defects in solids. Dissociation of oligomeric
proteins, like hemoglobin or disassembly (peptization)of actin and
microtubules could be also related with such big fluctuations.}

\begin{center}
{\large \medskip}

{\large \medskip5. \thinspace Mesoscopic mechanism of enzyme catalysis}
\end{center}

\smallskip

\textbf{The mechanism of enzyme catalysis is one of the most intriguing and
unresolved yet problems of molecular biology. It becomes clear, that it is
interrelated not only with a spatial, but as well with hierarchical
complicated dynamic properties of proteins (see book: ''The Fluctuating
Enzyme'' , Ed. by G.R.Welch, 1986).}

\smallskip

The [proteins + solvent] system should be considered as a cooperative one with
feedback links (Kaivarainen, 1985, 1992). Somogyi and Damjanovich (1986)
proposed a similar idea that collective excitations of protein structure are
interrelated with surrounded water molecules oscillations.

\smallskip

\textbf{The enzymatic reaction can be represented in accordance with our
dynamic model as a consequence of the following stages (K\"{a}iv\"{a}%
r\"{a}inen, 1989; [1] ).}

\textbf{\smallskip}

\textbf{The first stage: }%

\begin{equation}
(I)\qquad E^{b}+S\rightleftharpoons E^{b}S\tag{36}%
\end{equation}
- the collision of the substrate (S) with the open (b) state of the active
site [AS] cavity of enzyme (E).

The frequency of collisions between the enzyme and the substrate, whose
concentrations are $[C_{E}]$ and $[C_{S}]$, respectively, is expressed with
the known formula (Cantor and Schimmel, 1980):%

\begin{equation}
\nu_{\text{col}}=4\pi r_{0}(D_{E}+D_{S})\cdot N_{0}[C_{E}]\cdot[C_{S}]\tag{37}%
\end{equation}
where: $r_{0}=a_{E}+a_{S}$ is the sum of the enzyme's and substrate's
molecular radii; $N_{0}$ is the Avogadro number;%

\begin{equation}
D_{E}={\frac{kT}{6\pi\eta a_{E}}}\text{ \ \ and \ \ }D_{S}={\frac{kT}{6\pi\eta
a_{S}}}\tag{38}%
\end{equation}
- are the diffusion coefficients of the enzyme and substrate; k is the
Boltzmann constant; T is absolute temperature; $\eta$ is a solvent viscosity.

The probability of collision of (\textit{b}) state of the active site with
substrate is proportional to the ratio of the b-state outer cross section area
to the whole enzyme surface area:%

\begin{equation}
P_{b}={\frac{\emptyset^{b}}{\emptyset_{E}}\cdot F_{b}}\tag{39}%
\end{equation}

where: $F_{b}$ $=\frac{f_{b}}{f_{b}+f_{a}}\;$ is a fraction of time, the
active site [AS] spend in the open (b) - state.

So, the frequency of collision between the substrate and (\textit{b}) state of
the active site (AS) with account for (40), meaning the first stage of
reaction is:%

\begin{equation}
\nu_{col}^{b}=\nu_{col}\cdot P_{b}=k_{I}\tag{40}%
\end{equation}
\medskip\textbf{The second stage }of enzymatic reaction is a formation of the
primary enzyme-substrate complex:%

\begin{equation}
(II)\qquad E^{b}S\Leftrightarrow[E^{a^{*}}S]^{(1)}\tag{41}%
\end{equation}
It corresponds to transition of the active site cavity from the open
(\textit{b}) state to the closed (\textit{a}) one and stabilization the latter
state by a ligand.

The rate constant of the [$b\rightarrow a]$ transitions is derived with the
Stokes-Einstein and Eyring -Polany generalized equation (3):%

\begin{equation}
k_{\text{II}}^{b\rightarrow a^{*}}={\frac{kT}{\eta V}}\exp\left(
-{\frac{G_{st}^{b\rightarrow a}}{RT}}\right) \tag{42}%
\end{equation}

where: $\eta$ is the solvent viscosity; V is the effective volume of the
enzyme domain, whose diffusional reorientation accompanies the $(b\rightarrow
a)$ transition of the active site [AS].

The leftward shift of the [$a\Leftrightarrow b]$ equilibrium between two
states of the active site is concomitant with this stage of the reaction. It
reflects the \textit{principle of stabilized conformity,} related
to\textit{\ }AS domains movements that we have put forward in the previous section.

\smallskip

\textbf{The third stage: }%

\begin{equation}
(III)\;\;\;\;\;\text{ }[E^{a}S]\Leftrightarrow[E^{a^{*}}S^{*}]\tag{43}%
\end{equation}
represents the formation of the secondary specific complex. This process is
related to directed ligand diffusion into the active site cavity and the
dynamic adaptation of its geometry to the geometry of the active site.

\textbf{Here the Principle of Stabilized microscopic Conformity is realized,
when the AS change its geometry from (a) to }$(a^{*})$\textbf{\ without
domains reorientation.} The rate constant of this stage is determined by the
rate constant of substrate diffusion in the closed (\textit{a}) state of the
active site cavity. It is also expressed by \textbf{generalized kinetic
equation }(42), but with other values of variables:%

\begin{equation}
k_{S}^{1\rightarrow2^{*}}={\frac{1}{\tau_{s}^{*}}}\exp\left(  -{\frac
{G_{s}^{a}}{RT}}\right)  =k_{\text{III}}\tag{44}%
\end{equation}

where:
\begin{equation}
\tau_{s}=(v_{s}/k)\eta^{\text{in}}/T\tag{45}%
\end{equation}
is the correlation time of substrate of volume ($v_{s})$ in the (\textit{a})
state of the active site; $\eta^{\text{in}}$ is the internal effective
viscosity; $G_{S}^{a}$ is the activation energy of thermal fluctuations of
groups, representing small-scale dynamics (SS), which determine the directed
diffusion of a substrate in the active site [AS] closed cavity.

The directed character of ligand diffusion in AS can be determined by the
relaxation of a protein structure, due to perturbation of AS domains by
ligand. The relaxation changes were observed in many reactions of specific
protein-ligand complexes formation (K\"{a}iv\"{a}r\"{a}inen, 1985, 1989).

The complex formation [pair of domains forming the AS + substrate], followed
by these domain immobilization can be considered as an emergency of a new
enlarged protein primary effecton from two smaller ones, corresponding to less
independent AS domains or their compact ''nodes''.

\textbf{We assume that at this important stage, the waves B of the attacking
catalytic atoms $(\lambda_{B}^{c})$ and the attacked substrate atoms
$(\lambda_{B}^{S})$ start to overlap and interfere in such a way that
conditions for quantum-mechanical beats between them become possible.}

\textbf{Let us consider these conditions in more detail.}

\smallskip

According to classical statistics, every degree of freedom gets the energy,
which is equal to $kT/2$. This condition corresponds to harmonic approximation
when the mean potential and kinetic energies of particles are the same:%

\begin{equation}
V=T_{k}=mv^{2}/2\approx kT/2\tag{46}%
\end{equation}
The corresponding to such ideal case the de Broglie wave (wave B) length is
equal to:%

\begin{equation}
\lambda_{B}={\frac{h}{mv}}={\frac{h}{(\text{mkT})^{1/2}}}\tag{47}%
\end{equation}
For such condition the wave B length of proton at room temperature is nearly
2.5\AA, for a carbon atom it is about three times smaller and for oxygen -
four times as small.

In the latter two cases, the wave B lengths are comparable and even less than
the sizes of the atoms itself. So, their waves can not overlap and, beats
between them are not possible.

\textbf{However, in real condensed systems with quantum properties, including
the active sites of enzymes, the harmonic approximation is not valid because
}$(T_{k}/V)\ll1$\textbf{\ }$($\textbf{see Fig. 5\thinspace\thinspace
of\thinspace\thinspace[1]).\ Consequently, the kinetic energy of atoms of AS:}%

\begin{equation}
T_{k}\ll(1/2)kT\text{ \ \ \ and\ \ \ \ }\lambda_{B}\gg h/(mkT)^{1/2}\tag{47a}%
\end{equation}

\textbf{It must be taken into account that librations, in a general case are
presented by rotational-translational unharmonic oscillations of atoms and
molecules, but not by their rotational motions only (Coffey et al., 1984).}

\textbf{The length of waves B of atoms caused by a small translational
component of most probable impulse, related to librations is bigger than that
related to pure translations: }
\begin{equation}
\lbrack\lambda_{lb}=(h/P_{lb})]>[\lambda_{tr}=(h/P_{tr})]\tag{47b}%
\end{equation}

\textbf{Even in pure water the linear sizes of primary librational effectons
are several times bigger than that of translational effectons and the size of
one }$H_{2}O$\textbf{\ molecule (see Fig. 7 of [1] or Fig.4 of [2]).}

\textbf{In composition of the active site rigid core the librational waves B
of atoms can significantly exceed the sizes of the atoms themselves. In this
case their superposition, leading to quantum beats of waves B in the [active
site - substrate] complex, accelerating the enzyme reaction is quite possible.}

\textbf{\ In accordance to our model, periodic energy exchange resulting from
such beats occurs between waves B of the substrate and active site atoms.}

\smallskip

\textbf{The reaction }$\left[  S^{*}\rightarrow P\right]  $\textbf{,
accelerating by these quantum beats, is the next 4th stage of enzymatic
process - the chemical transformation of a substrate into product:}%

\begin{equation}
(IV)\qquad\;\;\;[E^{a^{*}}S^{*}]\rightarrow[E^{a^{*}}P]\tag{48}%
\end{equation}
The angular wave B frequency of the \textit{attacked} \textit{atoms of
substrate} with mass $m_{S}$ and the amplitude $A_{S}$ can be expressed by eq.(2.20):%

\begin{equation}
\omega^{S}=\hbar/2m_{S}A_{S}^{2}\tag{49}%
\end{equation}

The wave B frequency of the attacking catalytic atom (or a group of atoms) in
the active site is equal to:%

\begin{equation}
\omega^{\text{cat}}={\frac{\hbar}{2m_{c}A_{c}}}\tag{50}%
\end{equation}
\textit{The frequency of quantum beats }which appear between waves B of
catalytic and substrate atoms is:%

\begin{equation}
\omega^{*}=\omega^{\text{cat}}-\omega^{S}=\hbar\left(  {\frac{1}{2m_{c}A_{c}}%
}-{\frac{1}{2m_{S}A_{S}}}\right) \tag{51}%
\end{equation}

The corresponding energy of beats:%

\begin{equation}
E^{*}=E^{\text{cat}}-E^{S}=\hbar\omega^{*}\tag{52}%
\end{equation}
It is seen from these formulae that the smaller the wave B mass of the
catalytic atom $(m_{c})$ and its amplitude (A$_{c})$, the more frequently
these beats occur at constant parameters of substrate (m$_{S}$ $and$ A$_{S}%
)$.\textbf{\ The energy of beats is transmitted to the wave B of the attacked
substrate atom from the catalytic atom, accelerating the reaction.}

\smallskip

According to our model, the perturbations in the region of the active site are
accompanied by the appearance of phonons and acoustic deformons in a form of
small-scale dynamics of protein structure. They provide the signal
transmission in oligomeric proteins to the central cavity and other active
sites leading to allosteric effects.

\smallskip

It is known from the theory of oscillations (Grawford, 1973) that the effect
of beats is maximal, if the amplitudes of the interacting oscillators are equal:%

\begin{equation}
A_{c}^{2}\approx A_{S}^{2}\tag{53}%
\end{equation}
The [substrate $\rightarrow$ product] transformation can be considered as a
result of the substrate wave B transition from the main [S] state to excited
[P] state. The rate constant of such a reaction in the absence of the catalyst
($k^{S\rightarrow P})$ can be presented by the modified Eyring-Polany
formulae, leading from eq.(2.27 of [1]) at condition: $\exp(h\nu_{p}/kT)>>1$%

\begin{equation}
\nu_{A}^{S}=k^{S\rightarrow P}={\frac{E^{P}}{h}}\exp\left(  -{\frac
{E^{P}-E^{S}}{kT}}\right)  =\nu_{B}^{P}\exp\left[  -{\frac{h(\nu_{B}^{P}%
-\nu_{B}^{S})}{kT}}\right] \tag{54}%
\end{equation}

where: $E^{S}=h\nu^{S}\;$ and\ $E^{P}=h\nu^{P}$ are the main and excited -
transitional to product energetic states of substrate, correspondingly;
$\nu^{S}$ and $\nu^{P}$ are the substrate wave B frequencies in the main and
excited states, respectively.

\smallskip

\textbf{If catalyst is present,} which acts by the above described mechanism,
then the energy of the substrate E$^{S}$ is increased by the magnitude
$E^{\text{cat}}$ with the quantum beats frequency $\omega^{*}\;(51)$ and gets
equal to:%

\begin{equation}
E^{Sc}=E^{S}+E^{\text{cat}}\tag{55}%
\end{equation}
Substituting $E^{P}=h\nu_{B}^{P}\;$ and $\;E^{Sc}=h(\nu_{B}^{S}+\nu_{B}%
^{cat})$ in (54), we derive the rate constants for the catalytic reaction in
the moment of beats ($k^{Sc\rightarrow P}).$ This corresponds to the 4th stage
of enzymatic reaction:%

\begin{equation}
\text{(IV):\ \ \ }k^{Sc\rightarrow P}=\nu^{P}\exp\left[  -{\frac{h(\nu^{P}%
-\nu^{S}-\nu^{\text{cat}})}{kT}}\right]  =k_{\text{IV}}\tag{56}%
\end{equation}
where: $\nu^{P},\nu^{S}$ and $\nu^{c}$ are the most probable B wave
frequencies of the transition $[S\rightarrow P]$ state, of the substrate and
of the catalyst atoms, correspondingly.

\textbf{Hence, in the presence of the catalyst the coefficient of acceleration
(q) is equal to:}%

\begin{equation}
q_{\text{cat}}={\frac{k^{Sc\rightarrow P}}{k^{S\rightarrow P}}}=\exp\left(
{\frac{h\nu^{\text{cat}}}{kT}}\right) \tag{57}%
\end{equation}
For example, at%

\begin{equation}
h\nu^{c}/kT\approx10;\;\;\;\;\text{ }q_{\text{cat}}\text{ =}2.2\cdot
10^{4}\tag{58}%
\end{equation}
At room temperatures this condition corresponds to%

\[
E^{\text{cat}}=h\nu^{\text{cat}}\simeq6\text{ kcal/mole. }
\]
The beating acts, followed by transitions of a substrate molecule to activated
by catalyst excited state $(S\rightarrow Sc)$, can be accompanied by the
absorption of phonons or photons with the frequency $\omega^{*}\;(51) $. In
this case, the insolation of a [substrate - catalyst] system with ultrasound
or electromagnetic field of the frequency $\omega^{*} $ should strongly
accelerate the reaction when the resonance conditions are satisfied.

It is possible that the resonance effects of this type can account for the
experimentally revealed response of various biological systems to
electromagnetic field radiation with the frequency about $\,6\cdot10^{10}Hz$
$($Deviatkov et al., 1973). We have to point out that just such frequency is
close to the frequency of (a$\rightleftharpoons b)_{tr}$ transitions
excitation of primary translational water effectons. \textbf{The strict
correlation between the dynamics of water and that of biosystems should exist
on each hierarchic level of time and space.}

\smallskip

Changes in the volume, geometry and electronic properties of the substrate
molecule, resulting from its transition to the product, change its interaction
energy with the active site by the magnitude:%

\begin{equation}
\Delta E_{SP}^{a^{*}}=(E_{S}^{a^{*}}-E_{P}^{a^{*}})\tag{59}%
\end{equation}
It must destabilize the closed state of the active site and increase the
probability of its reverse [$a^{*}\rightarrow b^{*}]$ transition.

\textbf{Such transition promotes the last\ 5th stage of the catalytic cycle -
the dissociation of the enzyme-product complex:}%

\begin{equation}
(V)\;\;\;\qquad[E^{a^{*}}P]\Leftrightarrow[E_{P}^{b}P]\Leftrightarrow
E^{b}+P\tag{60}%
\end{equation}
The resulting rate constant of this stage, like stage (II), is described by
the generalized Stokes-Einstein and Eyring-Polany equation (42), but with
different activation energy $G_{st}^{a^{*}\rightarrow b}$ valid for the
[$a^{*}\rightarrow b]$ transition:%

\begin{equation}
k_{P}^{a^{*}\rightarrow b}={\frac{kT}{\eta V}}\exp\left(  -{\frac
{G_{st}^{a^{*}\rightarrow b}}{kT}}\right)  =1/\tau_{P}^{a^{*}\rightarrow
b}\tag{61}%
\end{equation}
If the lifetime of the $(a^{*})$ state is sufficiently long, then the
desorption of the product can occur irrespective of [$a^{*}\rightarrow b]$
transition, but with a longer characteristic time as a consequence of its
diffusion out of the active site's ''closed state''. The rate constant of this
process $(k_{p}^{*})$ is determined by small-scale dynamics. It practically
does not depend on solvent viscosity, but can grows up with rising
temperature, like at the 3d stage, described by (44):%

\begin{equation}
k_{P}^{*}={\frac{kT}{\eta_{a^{*}}^{\text{in}}\cdot v_{P}}}\exp\left(
-{\frac{G_{P}^{a^{*}}}{kT}}\right)  =1/\tau_{P}^{*}\tag{62}%
\end{equation}
where $\eta_{a^{*}}^{\text{in}}$ is active site interior viscosity in the
closed $(a^{*})$ state;\ $v_{P}$ is the effective volume of product molecules;
$P_{a^{*}}=\exp(-G_{P}^{a^{*}}/RT)$ is the probability of
small-scale,\ \textit{functionally important motions} necessary for the
desorption of the product from the $(a^{*})$ state of the active site;
$G_{p}^{a^{*}}$is the free energy of activation of such motions.

\smallskip

\textbf{Because the processes, described by the eq.(61) and (62), are
independent, the resulting product desorption rate constant is equal to:}%

\begin{equation}
k_{V}=k_{P}^{a^{*_{\rightarrow b}}}+k_{P}^{*}=1/\tau_{P}^{a^{*_{\rightarrow
b}}}+1/\tau_{P}^{*}\tag{63}%
\end{equation}

The characteristic time of this final stage of enzymatic reaction is:%

\begin{equation}
\tau_{V}=1/k_{V}=\frac{\tau_{P}^{a^{*_{\rightarrow b}}}\cdot\tau_{P}^{*}}%
{\tau_{P}^{a^{*_{\rightarrow b}}}+\tau_{P}^{*}}\tag{64}%
\end{equation}
This stage is accompanied by the relaxation of the perturbed AS domain and
remnant protein structure to the initial state.

After the whole reaction cycle is completed the enzyme gets ready for the next
cycle. The number of cycles (catalytic acts) in the majority of enzymes is
within the limits of $(10^{2}-10^{4})\,s^{-1}$. It means that the
[$a\rightleftharpoons b]$ pulsations of the active site cavities must occur
with higher frequency as far it is only one of the five stages of enzymatic
reaction cycle.

In experiments, where various sucrose concentrations were used at constant
temperature, the dependence of enzymatic catalysis rate on solvent viscosity
$(T/\eta)$ was demonstrated $($Gavish and Weber, 1979). The amendment for
changing the dielectric penetrability of the solvent by sucrose was taken into
account. There are reasons to consider stages (II) and/or (V) in the model
described above as the limiting ones of enzyme catalysis. According to
eqs.(43) and (61), these stages depend on $(T/\eta)$, indeed.

\smallskip

\textbf{The resulting rate constant of the enzyme reaction could be expressed
as the reciprocal sum of life times of all its separate stages (I-V):}%

\begin{equation}
k_{\text{res}}=1/\tau_{\text{res}}=1/(\tau_{I}+\tau_{\text{II}}+\tau
_{\text{III}}+\tau_{\text{IV}}+\tau_{V})\tag{65}%
\end{equation}

where

\begin{center}%
\begin{equation}
\tau_{I}=1/k_{I}={\frac{1}{\nu_{\text{col}}\cdot P_{b}}};\text{ \ \thinspace
\thinspace\thinspace\thinspace\thinspace\thinspace\thinspace}\tau_{\text{II}%
}=1/k_{\text{II}}={\frac{\eta V}{kT}}\exp\left(  {\frac{G_{st}^{b\rightarrow
a}}{RT}}\right)  ;\tag{65a}%
\end{equation}%

\begin{equation}
\tau_{\text{III}}=1/k_{\text{III}}={\frac{\eta^{\text{in}}v_{S}}{kT}}%
\exp\left(  {\frac{G_{SS}^{a}}{RT}}\right)  ;\tag{65b}%
\end{equation}

$\smallskip$

$\tau_{\text{IV}}=1/k_{\text{IV}}=\frac{1}{(\nu^{p})}\exp\left[  {\frac
{h(\nu^{s}-\nu^{p}-\nu^{c}}{kT}}\right]  ;$

$\smallskip$

$\tau_{V}$ \ corresponds to eq.(64).
\end{center}

\smallskip

\ The slowest stages of the reaction seem to be stages (II), (V), and stage
(III). The latter is dependent only on the small- scale dynamics in the region
of the active site.

Sometimes \textit{product desorption }goes on much more slowly than other
stages of the enzymatic process, i.e.
\[
k_{V}\ll k_{\text{III}}\ll k_{\text{II}}
\]
Then the resulting rate of the process $(k_{\text{res}})$ is represented by
its limiting stage (eq. 63):%

\begin{equation}
k_{\text{res}}\approx{\frac{kT}{\eta V}}\exp\left(  -{\frac{G_{st}%
^{a^{*}\rightarrow b}}{RT}}\right)  +k_{P}^{*}=1/\tau_{\text{res}}\tag{66}%
\end{equation}

The corresponding period of enzyme turnover: $\tau_{\text{res}}\approx\tau
_{V}\;\;(eq\,64)$. The internal medium viscosity ($\eta^{\text{in}}$) in the
protein regions, which are far from the periphery, is 2-3 orders higher than
the viscosity of a water-saline solvent (0.001 P) under standard conditions:
\[
(\eta^{\text{in}}/\eta)\ge10^{3}
\]
Therefore, the changes of sucrose concentration in the limits of 0-40\% at
constant temperature can not influence markedly internal small- scale dynamics
in proteins, its activation energy $(G^{a^{*}})$ and internal microviscosity
(K\"{a}iv\"{a}r\"{a}inen, 1989b). This fact was revealed using the spin-label
method. It is in accordance with viscosity dependencies of tryptophan
fluorescence quenching in proteins and model systems related to acrylamide
diffusion in protein matrix (Eftink and Hagaman, 1986). In the examples of
parvalbumin and ribonuclease $T_{1}$ it has been shown that the dynamics of
internal residues is practically insensitive to changing solvent viscosity by
glycerol over the range of 0.01 to 1 P.

It follows from the above data that the moderate changes in solvent viscosity
($\eta$) at constant temperature do not influence markedly the $k_{P}^{*}$
value in eq.(66).

Therefore, the isothermal dependencies of k$_{\text{res}}$ on $(T/\eta)_{T}$
with changing sucrose or glycerol concentration must represent straight lines
with the slope:%

\begin{equation}
tg\alpha={\frac{\Delta k_{\text{res}}}{\Delta(T/\eta)_{T}}}={\frac{k}{V}}%
\exp\left(  -{\frac{G_{st}^{a^{*}\rightarrow b}}{RT}}\right)  _{T}\tag{67}%
\end{equation}
The interception of isotherms at extrapolation to $(T/\eta\rightarrow0)$
yields (62):%

\begin{equation}
\left(  k_{\text{res}}\right)  _{(T/\eta)\rightarrow0}=k_{P}^{*}={\frac
{kT}{\eta_{a^{*}}^{\text{in}}v_{P}}}\exp\left(  -{\frac{G^{a^{*}}}{RT}}\right)
\tag{68}%
\end{equation}
The volume of the Brownian particle (V) in eq.(67) corresponds to the
effective volume of one of the domains, which reorientation is responsible for
(a$\rightleftharpoons b)$ transitions of the enzyme active site.

Under conditions when the activation energy G$_{st}^{a^{*}\rightarrow b}$
weakly depends on temperature, it is possible to investigate the temperature
dependence of the effective volume V, using eq.(67), analyzing a slopes of set
of isotherms (67).

Our model predicts the increasing of V with temperature rising. This reflects
the dumping of the large-scale dynamics of proteins due to water clusters
melting and enhancement the Van der Waals interactions between protein domains
and subunits (K\"{a}iv\"{a}r\"{a}inen, 1985;\thinspace1989b,\ K\"{a}%
iv\"{a}r\"{a}inen et al., 1993). The contribution of the small-scale dynamics
$(k^{*})$ to $k_{\text{res}}$ must grow due to its thermoactivation and the
decrease in $\eta^{\text{in}}$ and $G^{a^{*}}(eq.32)$.

\smallskip

\textbf{The diffusion trajectory of ligands, substrates and products of enzyme
reactions in ''closed'' (a) states of active sites is probably determined by
the spatial gradient of minimum wave B length (maximum impulses) values of
atoms, forming the active site cavity.}

\textbf{\ }We suppose, that \thinspace\textit{functionally important motions
(FIM)}, introduced by H. Frauenfelder et al., (1985, 1988), are determined by
specific geometry of the impulse space characterizing the distribution of
small-scale dynamics of domains in the region of \ protein's active site.

\textbf{The analysis of the impulse distribution in the active site area and
energy of quantum beats between de Broglie waves of the atoms of substrate and
active sites, modulated by solvent-dependent large-scale dynamics, should lead
to complete understanding of the physical background of enzyme catalysis.}

\medskip

\begin{center}
{\large \smallskip}

{\large 6. The mechanism of ATP hydrolysis energy utilization in muscle
contraction }

{\large and protein polymerization}
\end{center}

\smallskip

\textbf{A great number of biochemical reactions are endothermic, i.e. they
need additional thermal energy in contrast to exothermic ones. The most
universal and common source of this additional energy is a reaction of
adenosinetriphosphate (ATP) hydrolysis:}%

\begin{equation}
\text{ATP }
\begin{array}
[c]{l}%
k_{1}\\
\Leftrightarrow\\
k_{-1}%
\end{array}
\text{ ADP + P}\tag{69}%
\end{equation}
The reaction products are adenosinediphosphate (ADP) and inorganic phosphate (P).

The equilibrium constant of the reaction depends on the concentration of the
substrate [ATP] and products [ADP] and [P] like:%

\begin{equation}
K={\frac{k_{1}}{k_{-1}}}={\frac{[\text{ADP]}\cdot\text{[P]}}{[\text{ATP]}}%
}\tag{70}%
\end{equation}
The equilibrium constant and temperature determine the reaction free energy change:%

\begin{equation}
\Delta G=-RT\ln K=\Delta H-T\Delta S\tag{71}%
\end{equation}
where: $\Delta$H and $\Delta$S are changes in enthalpy and entropy, respectively.

\smallskip

\textbf{Under the real conditions in cell the reaction of ATP hydrolysis is
highly favorable energetically as is accompanied by strong free energy
decrease: }$\Delta G=-(11\div13)$\textbf{\ kcal/M.}

\textbf{It follows from (71) that }$\Delta G<0$\textbf{, when}%

\begin{equation}
T\Delta S>\Delta H\tag{72}%
\end{equation}
\textbf{and the entropy and enthalpy changes are positive }$(\Delta
S>0$\textbf{\ \ and }$\;\Delta H>0).$\textbf{\ However, the specific molecular
mechanism of these changes in different biochemical reactions, including
muscle contraction, remains unclear.}

\textbf{Acceleration of actin polymerization and tubulin self-assembly to the
microtubules as a result of the ATP and nucleotide GTP splitting,
respectively, is still obscure as well.}

\smallskip

\textbf{Using our model of water-macromolecule interaction [6], we can explain
these processes by the ''melting'' of the water clusters - librational
effectons in cavities between neighboring domains and subunits of proteins.
This melting is induced by absorption of energy of ATP or GTP hydrolysis and
represents }$[lb/tr]$\textbf{\ conversion of primary librational effectons to
translational ones. It leads to the partial dehydration and rapprochement of
domains and subunits. The concomitant transition of interdomain/subunit
cavities from the ''open'' B-state to the ''closed'' A-state should be
accompanied by decreasing of linear dimensions of a macromolecule. This
process is usually reversible and responsible for the large-scale dynamics.}

\textbf{In the case when disjoining clusterphilic interactions that shift the
}$\left[  \mathbf{A\Leftrightarrow B}\right]  $\textbf{\ equilibrium to the
right are stronger than Van der Waals interactions stabilizing A-state, the
expansion of the macromolecule can induce a mechanical ''pushing'' force.}

{\large In accordance to our model, this ''swelling driving force'' is
responsible for shifting of myosin ''heads'' as respect to the actin filaments
and muscle contraction.}

\textbf{\ Such FIRST relaxation ''swelling working step''} is accompanied by
dissociation of products of ATP hydrolysis from the active sites of myosin
heads (heavy meromyosin).

\textbf{The SECOND stage of reaction,} the dissociation of the complex:
[myosin ''head'' + actin], is related to the absorption of ATP at the myosin
active site. At this stage the $A\Leftrightarrow B$ equilibrium between the
heavy meromyosin conformers is strongly shifted to the right, i.e. to an
expanded form of the protein.

\textbf{The THIRD stage is represented by the ATP hydrolysis, (ATP
}$\rightarrow$ \textbf{ADP + P) and expelling of (P) from the active site. The
concomitant local enthalpy and entropy jump leads to the melting of the water
clusters in the cavities, }$\mathbf{B}\rightarrow\mathbf{A}$
\textbf{transitions and the contraction of free meromyosin heads.}

\textit{The energy of the clusterphilic interaction at this stage is
accumulated in myosin like in a squeezed spring.}

\smallskip

\textbf{After this 3d stage is over the complex [myosin head + actin] forms again.}

\smallskip

\textbf{We assume here that the interaction between myosin head and actin
induces the releasing of the product (ADP) from myosin active site.\ It is
important to stress that the driving force of \ ''swelling working
stage}$":[A\rightarrow B]$\textbf{\ transition of myosin cavities - is
represented by our clusterphilic interactions (see Section 13.3 of [1] and
paper [5]).}

\textbf{A repetition of such a cycle results in the relative shift of myosin
filaments with respect to actin ones and finally in muscle contraction.}

\smallskip

The mechanism proposed does not need the hypothesis of Davydov's soliton
propagation (Davydov, 1984) along a myosin macromolecule. It seems that this
nondissipative process scarcely takes place in strongly fluctuating biological
systems. Soliton model does not take into account the real mesoscopic
structure of macromolecules and their interaction with water as well.

\smallskip

\ \textbf{Polymerization of actin, tubulin and other globular proteins
composing cytoplasmic and extracell filaments due to hydrophobic interaction
can be accelerated as a result of their selected dehydration due to local
temperature jumps in mesoscopic volumes where the ATP and GTP hydrolysis takes place.}

\ The [assembly $\Leftrightarrow$ disassembly] equilibrium is shifted as a
result of such mesophase transition to the left in the case when $\left[
protein-protein\right]  $ interface Van-der-Waals interactions are stronger
than a clusterphilic one. The latter is mediated by librational water effecton
stabilization in interdomain or intersubunit cavities.

\textbf{It looks that the clusterphilic interactions play an extremely
important role on mesoscale in the self-organization and dynamics of
biological systems.}

\begin{center}
\smallskip

{\large 7. Water activity as a regulative factor in the intra- and inter-cell processes}
\end{center}

\smallskip

\textbf{Three most important factors can be responsible for the spatial
processes in living cells:}

\textbf{1.~Self-organization of supramolecular systems in the form of
membranes, oligomeric proteins and filaments. Such processes can be mediated
by water-dependent hydrophilic, hydrophobic and introduced by us clusterphilic
interactions [1, 5];}

\textbf{2.~Compartmentalization of cell volume by semipermeable lipid- bilayer
membranes and due to different cell's organelles formation;}

\textbf{3.~Changes in the volumes of different cell compartments by osmotic
process correlated in space and time. These changes are dependent on water
activity regulated, in turn, mainly by [assembly}$\rightleftharpoons
$\textbf{\thinspace disassembly] equilibrium shift of microtubules and actin's filaments.}

\smallskip

Disassembly of filaments leads to water activity decreasing due to increasing
the fraction of \textbf{vicinal water, representing a developed system of
enlarged primary librational effectons} near the surface of proteins (see
section 13.5 of [1] and [5]). The thickness of vicinal water layer is about 50
\AA , depending on temperature and mobility of intra-cell components.

The vicinal water with more ordered and cooperative structure than that of
bulk water represents the dominant fraction of intra-cell water. A lot of
experimental evidences, pointing to important role of vicinal water in cell
physiology were presented in reviews of Drost-Hansen and Singleton (1992) and
Clegg and Drost-Hansen (1991).

\smallskip

The dynamic equilibrium: $\{$ I $\Leftrightarrow$ II $\Leftrightarrow$ III
$\}$ between three stages of macromolecular self-organization, discussed in
Section 13.5 of [1] (Table 2) and in paper [5], has to play an important role
in biosystems: blood, lymph as well as inter- and intra-cell media. This
equilibrium is dependent on the water activity (inorganic ions, pH),
temperature, concentration and surface properties of macromolecules.

Large-scale protein dynamics, decreasing the fraction of vicinal water
(K\"{a}iv\"{a}r\"{a}inen, 1986, K\"{a}iv\"{a}r\"{a}inen et al., 1990) is
dependent on the protein's active site ligand state. These factors may play a
regulative role in [\textit{coagulation }$\Leftrightarrow$
\textit{peptization}] and [\textit{gel} $\Leftrightarrow$ \textit{sol}]
transitions in the cytoplasm of mobile cells, necessary for their migration.

A lot of spatial cellular processes such as the increase or decrease in the
length of microtubules or actin filaments are dependent also on their
\textit{self-assembly }from corresponding subunits $(\alpha,\beta$ tubulins
and actin).

The self-assembly of such superpolymers is dependent on the [association
$(A)\Leftrightarrow$ dissociation (B)] equilibrium constant
$(K_{A\Leftrightarrow B}=K_{B\Leftrightarrow A}^{-1})$. In turn, this constant
is dependent on water activity $(a_{H_{2}O})$, as was shown earlier
$(eq.13.11a$ and 13.12 of [1] and [5]).

\textit{The double helix of actin filaments}, responsible for the spatial
organization and cell's shape dynamics, is composed of the monomers of
globular protein - actin (MM 42.000). The rate of actin filament
polymerization or depolymerization, responsible for cells shape adaptation to
environment, is very high and strongly depends on ionic strength
(concentration of $NaCl$, $Ca^{++},\;Mg^{++})$. For example, the increasing of
$NaCl$ concentration and corresponding decreasing of $a_{H_{2}O}$ stimulate
the actin polymerization. The same is true of $\alpha$ and $\beta$ tubulin
polymerization in the form of microtubules.

The activity of water in cells and cell compartments can be regulated by
$\left[  Na^{+}-K^{+}\right]  \;ATP-$dependent pumps. Even the equal
concentrations of $Na^{+}$ and $K^{+}$ decrease water activity $a_{H_{2}O}$
differently due to their different interaction with bulk and, especially with
ordered vicinal water (Wiggins 1971, 1973).

Regulation of pH by proton pumps, incorporated in membranes, also can be of
great importance for intra-cell $a_{H_{2}O}$ changing.

Cell division is strongly correlated with dynamic equilibrium: [assembly
$\Leftrightarrow$ disassembly] of microtubules of centrioles.. Inhibition of
tubulin subunits dissociation (disassembly) by \textit{addition of }$D_{2}O$,
or \textit{stimulation} this process by \textit{decreasing temperature or
increasing hydrostatic pressure} stops cell mitosis - division (Alberts et
al., 1983).

The above mentioned factors enable to affect the $A\Leftrightarrow B$
equilibrium of cavity between $\alpha\,\,and\,\,\beta$ tubulins, composing
microtubules. These factors action confirm our hypothesis, that microtubules
assembly are mediated by clusterphilic interaction (see section 17.5 of [1]
and paper [5]).

The decrease in temperature and increase in intra-microtubules pressure lead
to the increased dimensions of librational water effectons, clustrons and
finally this induce disassembly of microtubules.

Microtubules are responsible for the coordination of intra-cell space
organization and movements, including chromosome movement at the mitotic
cycle, coordinated by centrioles.

The communication between different cells by means of channels can regulate
the ionic concentration and correspondent $a_{H_{2}O}$ gradients in the embryo.

\textbf{In accordance with our hypothesis, the gradient of water activity,
regulated by change of vicinal water fraction in different compartments of
cell can play a role of so-called morphogenic factor necessary for
differentiation of embryo cells.}

\medskip

\begin{center}
{\large 8. Water and cancer}
\end{center}

\smallskip

\textbf{We put forward a hypothesis that unlimited cancer cell division is
related to partial disassembly of cytoskeleton's actin-like filaments}
\textbf{due to some genetically controlled mistakes in biosynthesis and
increasing the osmotic diffusion of water into transformed cell.}

Decreasing of the intra-cell concentration of any types of ions $(Na^{+}%
,\;K^{+},\;H^{+}$, $Mg^{2+}\,$etc.), as the result of corresponding ionic pump
destruction, incorporated in biomembranes, also may lead to disassembly of filaments.

The shift of equilibrium: [assembly $\Leftrightarrow$ disassembly] of
microtubules (MTs) and actin filaments to the right increases the amount of
intra-cell water, involved in hydration shells of protein and decreases water
activity. As a consequence of concomitant osmotic process, cells tend to swell
and acquire a ball-like shape. The number of direct contacts between
transformed cells decrease and the water activity in the intercell space
increases also.

\textbf{We suppose that certain decline in the external inter-cell water
activity could be a triggering signal for the inhibition of normal cell
division. The shape of normal cells under control of cell's filament is a
specific one, providing good dense intercell contacts with limited amount of
water, in contrast to transformed cells.}

{\large If this idea is true, the absence of contact inhibition in the case of
cancer cells is a result of insufficient decreasing of intercell water
activity due to loose [cell-cell] contacts.}

\textbf{\smallskip}

\textbf{If our model of cancer emergency is correct, then the problem of tumor
inhibition is related to the problem of inter - and intra-cell water activity
regulation by means of chemical and physical factors.}

\textbf{\smallskip}

\textbf{Another approach for cancer healing we can propose here is the IR
laser treatment of transformed cells with IR\ photons frequencies, stimulating
superdeformons excitation and collective disassembly of MTs in composition of
centrioles. This will prevent cells division and should have a good
therapeutic effect.\ This approach is based on assumption that stability of
MTs in transformed cells is weaker and/or resonant frequency of their
superdeformons excitation differs from that of normal cells.}

\bigskip

\begin{center}
====================================================================\bigskip

\textbf{REFERENCES}
\end{center}

\begin{quotation}
\textbf{\medskip}

\textbf{Alberts B., Bray D., Lewis J., Ruff M., Roberts K. and Watson J.D.
Molecular Biology of Cell. Chapter 10. Garland Publishing, Inc.New York,
London, 1983.}

\textbf{Antonchenko V.Ya. Physics of water. Naukova dumka, Kiev, 1986.}

\textbf{Cantor C.R., Schimmel P.R. Biophysical Chemistry. W.H.Freemen and
Company, San Francisco, 1980.}

\textbf{Clegg J. S. On the physical properties and potential roles of
intracellular water. Proc.NATO Adv.Res.Work Shop. 1985.}

\textbf{Clegg J.S. and Drost-Hansen W. On the biochemistry and cell physiology
of water. In: Hochachka and Mommsen (eds.). Biochemistry and molecular biology
of fishes. Elsevier Science Publ. vol.1, Ch.1, pp.1-23, 1991.}

\textbf{Coffey W., Evans M., Grigolini P. Molecular diffusion and spectra.
A.Wiley Interscience Publication, N.Y., Chichester, Toronto, 1984.}

\textbf{Davydov A.S. Solitons in molecular systems. Phys. Scripta,
}$1979,\,20,\,387-394$\textbf{.}

\textbf{Davydov A.S. Solitons in molecular systems. Naukova dumka, Kiev, 1984
(in Russian).}

\textbf{Del Giudice E., Dogulia S., Milani M. and Vitello G. A quantum field
theoretical approach to the collective behaviour of biological systems.
Nuclear Physics\ }$1985,$\textbf{\ }$B251[FS13],375-400$\textbf{.}

\textbf{Drost-Hansen W. In: Colloid and Interface Science. Ed. Kerker M.
Academic Press, New York, 1976, p.267.}

\textbf{Drost-Hansen W., Singleton J. Lin. Our aqueous heritage: evidence for
vicinal water in cells. In: Fundamentals of Medical Cell Biology, v.3A,
Chemisrty of the living cell, JAI\ Press Inc.,1992, p.157-180.}

\textbf{Eftink M.R., Hagaman K.A. Biophys.Chem. 1986, 25, 277.}

\textbf{Einstein A. Collection of works. Nauka, Moscow, 1965 (in Russian).}

\textbf{Eisenberg D., Kauzmann W. The structure and properties of water.
Oxford University Press, Oxford, 1969.}

\textbf{Gavish B., Weber M. Viscosity-dependent structural fluctuations in
enzyme catalysis. Biochemistry 18\thinspace(1979)\thinspace1269.}

\textbf{Gavish B. in book: The fluctuating enzyme. Ed. by G.R.Welch. Wiley-
Interscience Publication, 1986, p.264-339.}

\textbf{Grawford F.S. Waves. Berkley Physics Course. Vol.3. McGraw- Hill Book
Co., N.Y., 1973.}

\textbf{Grundler W. and Keilmann F. Sharp resonance in Yeast growth proved
nonthermal sensitivity to microwaves. Phys.Rev.Letts., }$1983,51,1214-1216$\textbf{.}

\textbf{K\"{a}iv\"{a}r\"{a}inen A.I. Solvent-dependent flexibility of proteins
and principles of their function. D.Reidel Publ.Co., Dordrecht, Boston,
Lancaster, 1985,\thinspace pp.290.}

\textbf{K\"{a}iv\"{a}r\"{a}inen A.I. The noncontact interaction between
macromolecules revealed by modified spin-label method. Biofizika
(USSR}$)\;1987,\,32,\,536$\textbf{.}

\textbf{K\"{a}iv\"{a}r\"{a}inen A.I. Thermodynamic analysis of the system:
water-ions-macromolecules. Biofizika (USSR}$),\,1988,\,33,\,549$\textbf{.}

\textbf{K\"{a}iv\"{a}r\"{a}inen A.I. Theory of condensed state as a
hierarchical system of quasiparticles formed by phonons and three-dimensional
de Broglie waves of molecules. Application of theory to thermodynamics of
water and ice. J.Mol.Liq. }$1989a,\,41,\,53-60$\textbf{.}

\textbf{K\"{a}iv\"{a}r\"{a}inen A.I. Mesoscopic theory of matter and its
interaction with light. Principles of selforganization in ice, water and
biosystems. University of Turku, Finland\thinspace1992, pp.275.}

\textbf{Kaivarainen A. Dynamic model of wave-particle duality and Grand
unification. University of Joensuu, Finland 1993. pp.118.}

\textbf{Kaivarainen A. Mesoscopic model of elementary act of perception and
braining. Abstracts of conference: Toward a Science of Consciousness 1996,
p.74. Tucson, USA.}

\textbf{K\"{a}iv\"{a}r\"{a}inen A., Fradkova L., Korpela T. Separate
contributions of large- and small-scale dynamics to the heat capacity of
proteins. A new viscosity approach. Acta Chem.Scand. }$1993,47,456-460$\textbf{.}

\textbf{Karplus M., McCammon J.A. Scientific American, April 1986, p.42.}

\textbf{Koshland D.E. J.Theoret.Biol. 2(1962)75.}

\textbf{Kovacs A.L. Hierarcical processes in biological systems. Math. Comput.
Modelling. }$1990,\,14,\,674-679$\textbf{.}

\textbf{Lumry R. and Gregory R.B. Free-energy managment in protein reactions:
concepts, complications and compensations. In book: The fluctuating enzyme. A
Wiley-Interscience publication. 1986, p. 341- 368.}

\textbf{Schulz G.E., Schirmer R.H. Principles of protein structure.
Springer-Verlag, New York, 1979.}

\textbf{Somogyi B. and Damjanovich S. In book: The fluctuating enzyme. Ed. by
G.R.Welch. A Wiley-InterScience Publication. 1986, 341-368.}

\textbf{Wiggins P.M. Thermal anomalies in ion distribution in rat kidney
slices and in a model system. Clin. Exp. Pharmacol. Physiol. 1972, 2, 171-176.}
\end{quotation}

\begin{center}
\end{center}
\end{document}